\documentclass[a4paper,11pt]{article}
\pdfoutput=1 % if your are submitting a pdflatex (i.e. if you have
\usepackage{jcappub} % for details on the use of the package, please
\usepackage[T1]{fontenc} % if needed

\title{Signals induced by charge-trapping in EDELWEISS FID detectors: analytical modeling and applications.}
\collaboration{The EDELWEISS Collaboration}

\author[a,1]{Q.~Arnaud\note{Corresponding author. Now at Queen's University, Kingston, Canada}}  %\fnref{fn1}
\author[b]{E.~Armengaud}
\author[a]{C.~Augier}
\author[c]{A.~Beno\^{i}t}
\author[d]{L.~Berg\'{e}}
\author[a]{J.~Billard}
\author[e,f]{J.~Bl\"{u}mer}
\author[b]{T. de~Boissi\`{e}re}
\author[d,e]{A.~Broniatowski}
\author[c]{P.~Camus}
\author[a]{A.~Cazes}
\author[d]{M.~Chapellier}
\author[a]{F.~Charlieux}
\author[d]{L.~Dumoulin}
\author[f]{K.~Eitel}
\author[e]{N.~Foerster}
\author[b]{N.~Fourches}
\author[a]{J.~Gascon}
\author[d]{A.~Giuliani}
\author[b]{M.~Gros}
\author[f]{L.~Hehn} 
\author[e]{G.~Heuermann}
\author[a]{M. De~J\'{e}sus}
\author[g]{Y.~Jin}
\author[a]{A.~Juillard}
\author[h]{M.~Kleifges}
\author[f]{V.~Kozlov}
\author[i]{H.~Kraus}
\author[a,e]{C.~K\'{e}f\'{e}lian}
\author[j]{V.~A.~Kudryavtsev}
\author[d]{H.~Le-Sueur}
\author[d]{S.~Marnieros}
\author[b]{X.-F.~Navick}
\author[b]{C.~Nones}
\author[d]{E.~Olivieri}
\author[k]{P.~Pari}
\author[b]{B.~Paul}
\author[d,2]{M.-C.~Piro \note{Now at Rensselaer Polytechnic Institute, Troy, NY, USA}}% \fnref{fn2}
\author[d]{D.~Poda}
\author[a]{E.~Queguiner}
\author[l]{S.~Rozov}
\author[a]{V.~Sanglard}
\author[f,3]{B.~Schmidt \note{Now at Lawrence Berkeley National Laboratory, Berkeley, CA, USA}} %\fnref{fn3}
\author[e]{S.~Scorza}
\author[f]{B.~Siebenborn}
\author[h]{D.~Tcherniakhovski}
\author[a]{L.~Vagneron}
\author[h]{M.~Weber}
\author[l]{E.~Yakushev}

\affiliation[a]{Univ Lyon, Universit\'{e} Claude Bernard Lyon 1,  CNRS-IN2P3, Institut de Physique Nucl\'{e}aire de Lyon, F-69622, Villeurbanne, France}
\affiliation[b]{CEA Saclay, DSM/IRFU, 91191 Gif-sur-Yvette Cedex, France}
\affiliation[c]{Institut N\'{e}el, CNRS/UJF, 25 rue des Martyrs, BP 166, 38042 Grenoble, France}
\affiliation[d]{CSNSM, Univ. Paris-Sud, CNRS/IN2P3, Universit\'{e} Paris-Saclay, 91405 Orsay, France}
\affiliation[e]{Karlsruher Institut f\"{u}r Technologie, Institut f\"{u}r Experimentelle Kernphysik, Gaedestr. 1, 76128 Karlsruhe, Germany}
\affiliation[f]{Karlsruher Institut f\"{u}r Technologie, Institut f\"{u}r Kernphysik, Postfach 3640, 76021 Karlsruhe, Germany}
\affiliation[g]{Laboratoire de Photonique et de Nanostructures, CNRS, Route de Nozay, 91460 Marcoussis, France}
\affiliation[h]{Karlsruher Institut f\"{u}r Technologie, Institut f\"{u}r Prozessdatenverarbeitung und Elektronik, Postfach 3640, 76021 Karlsruhe, Germany}
\affiliation[i]{University of Oxford, Department of Physics, Keble Road, Oxford OX1 3RH, UK}
\affiliation[j]{University of Sheffield, Department of Physics and Astronomy, Sheffield, S3 7RH, UK}
\affiliation[k]{CEA Saclay, DSM/IRAMIS, 91191 Gif-sur-Yvette Cedex, France}
\affiliation[l]{JINR, Laboratory of Nuclear Problems, Joliot-Curie 6, 141980 Dubna, Moscow Region, Russian Federation}

\emailAdd{q.arnaud@queensu.ca}
\abstract{
The EDELWEISS-III direct dark matter search experiment uses cryogenic HP-Ge detectors Fully covered with Inter-Digitized electrodes (FID). 
They are operated at low fields ($<1\;\mathrm{V/cm}$), and as a consequence charge-carrier trapping significantly 
affects both the ionization and heat energy measurements. 
This paper describes an analytical model of the signals induced by trapped charges in FID detectors based on the Shockley-Ramo theorem. 
It is used to demonstrate that veto electrodes, initially designed for the sole purpose of surface event rejection, 
can be used to provide a sensitivity to the depth of the energy deposits, characterize the trapping in the crystals, 
perform heat and ionization energy corrections and improve the ionization baseline resolutions. 
These procedures are applied successfully to actual data.
}

\begin{document}
\maketitle
\flushbottom

\section{Introduction}

\label{sec:intro}
The EDELWEISS-III experiment performs direct dark matter searches using HP-Ge bolometers operating at a cryogenic temperature of 18 mK in order to detect the low energy recoils $\mathcal{O}(10\;\mathrm{keV})$ expected from WIMP elastic scattering on target Ge nuclei~\cite{EDWIIIlowmass}. 
The sensitivity to WIMP-nucleon cross sections below $10^{-9}\;\mathrm{pb}$ requires powerful 
means to reduce and discriminate backgrounds. 
To this end, a simultaneous measurement of heat and ionization is performed to discriminate nuclear recoils 
induced by WIMPs from electronic recoils originating from $\gamma$- and $\beta$-rays \cite{shutt}. 
The latter are known \cite{IDdesign} to be potentially misidentified as nuclear recoils if most of the energy is deposited within a few $\mathrm{\mu m}$ from the surface of the crystals, where defects in the lattice and charge diffusion result in a biased ionization yield measurement. 
An active rejection of surface events is therefore mandatory and well achieved by the specific design of Fully Inter-Digitized (FID) detectors ~\cite{IDdesign}. 
The dimensions of the cylindrically shaped crystals are 70 mm diameter and 40 mm height. 
\begin{figure}[hbt] 
\centering 
\includegraphics[width=0.6\textwidth,keepaspectratio,clip,trim={5cm 0 0 0}]{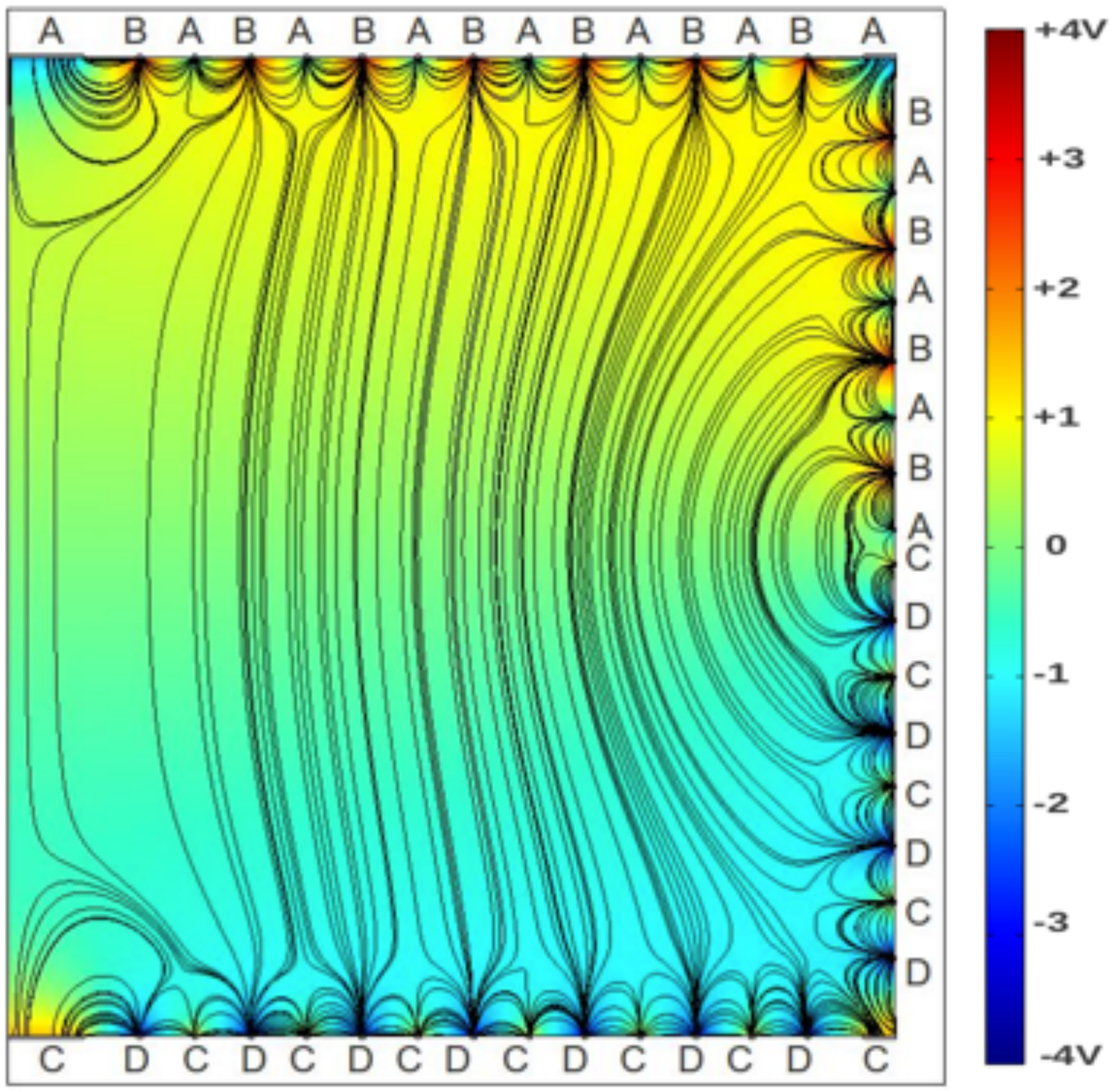}
\caption{\textit{Electric potential map and resulting field lines for a radial cross section of an FID detector. }}
 \label{fig1}
 \end{figure}
Concentric aluminum electrodes are evaporated over the entire Ge absorber surface.
They are alternatively bonded to define four ionization channels labelled A, B, C and D, as shown in Fig.~\ref{fig1}. 
Two sets of electrodes are biased (A=-1.5 V, B=+4 V) on one side and (C=+1.5 V, D=-4 V) 
on the other side, leading to an electric field structure as illustrated in Fig.~\ref{fig1}. 
Such low biases are applied to preserve the electronic/nuclear recoil discrimination power from being overwhelmed 
by the Neganov-Luke effect \cite{luke,neganov} 
whereby the heat signal is the sum of the original recoil energy plus a contribution proportional to 
the ionization signal and to the applied bias.
Following an energy deposit in the bulk, created electrons and holes drift to the electrodes with the highest bias, B and D respectively (hence called ``fiducial electrodes'') whereas for surface events charge collection is shared between one fiducial electrode and one so-called veto electrode (either B/A or D/C).  
The principle of the fiducial selection is thus to require an equal charge of opposite sign on fiducial electrodes and no charge collected on the veto electrodes A and C. 
This selection rejects events at depths below the surface comparable to the electrode spacing of 2 mm. 
Experimentally, the dispersion of the veto ionization measurement of fiducial events increases with energy,
as observed in Fig.~\ref{fig:coupfid}, reaching values that largely exceed the baseline resolutions.
These non-gaussian energy dispersions must be taken into account while performing fiducial cuts to avoid an unnecessary selection efficiency loss. 
It is therefore important to understand the origin of the signals on veto electrodes in these events where in principle
they should not collect any net charge.

We present an analytical model whereby this effect is understood as a consequence of charge carrier trapping in the fiducial volume\footnote{This approach is similar to the one of \protect{\cite{CdZnTeAnalytical}} for coplanar grid detectors sensitive to single-polarity 
charges~\protect{\cite{lukefromme,lukefromalexB}}.}. 
Charge trapping is expected to be significant \cite{alexJLTDNIM} due to the operating low field in the bulk (0.625 V/cm)~\cite{AlexBLTD}, 
affecting both the ionization and heat measurements through incomplete charge collection and Neganov-Luke effect, respectively. 
According to this model, charge-carriers trapped in the bulk induce signals on all the electrodes that are sufficient to account for the observed dispersions on veto electrodes for fiducial events. 
Initially designed for surface event rejection, 
the veto electrode signals contain information on energy deposit localization and bulk trapping properties. 
Thus the new applications investigated here lead to bulk trapping characterization
and correction procedures to improve the resolution of heat and ionization signals, at both low and high energy.

 \begin{figure}[htb] 
\centering 
\includegraphics[width=0.44\textwidth,keepaspectratio]{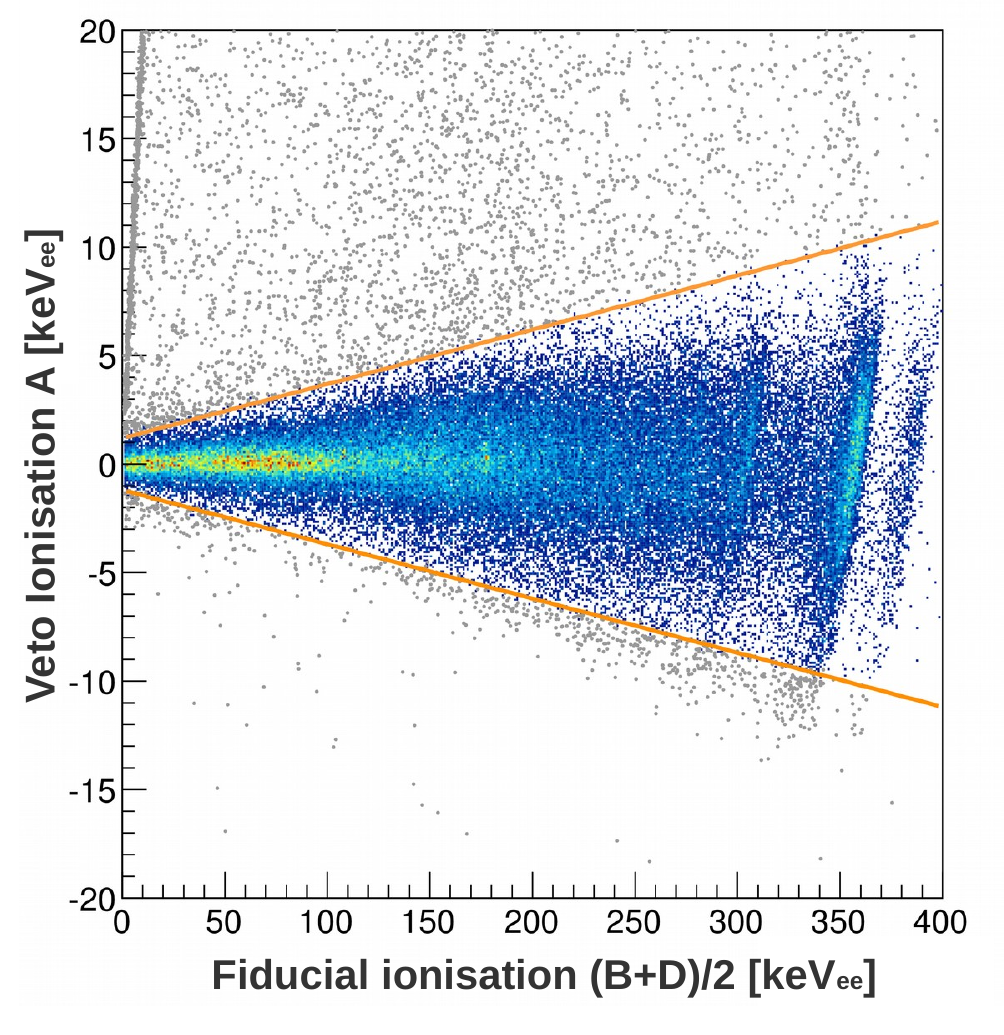}
\caption{\textit{Example of fiducial cut (orange solid lines) on the veto A. Fiducial events are represented in color dots whereas non fiducial events are shown in gray dots.}}
 \label{fig:coupfid}
 \end{figure}
 
\section{Weighting fields in FID detectors}

\begin{figure}[t] 
\centering 
\includegraphics[width=0.7\textwidth,keepaspectratio]{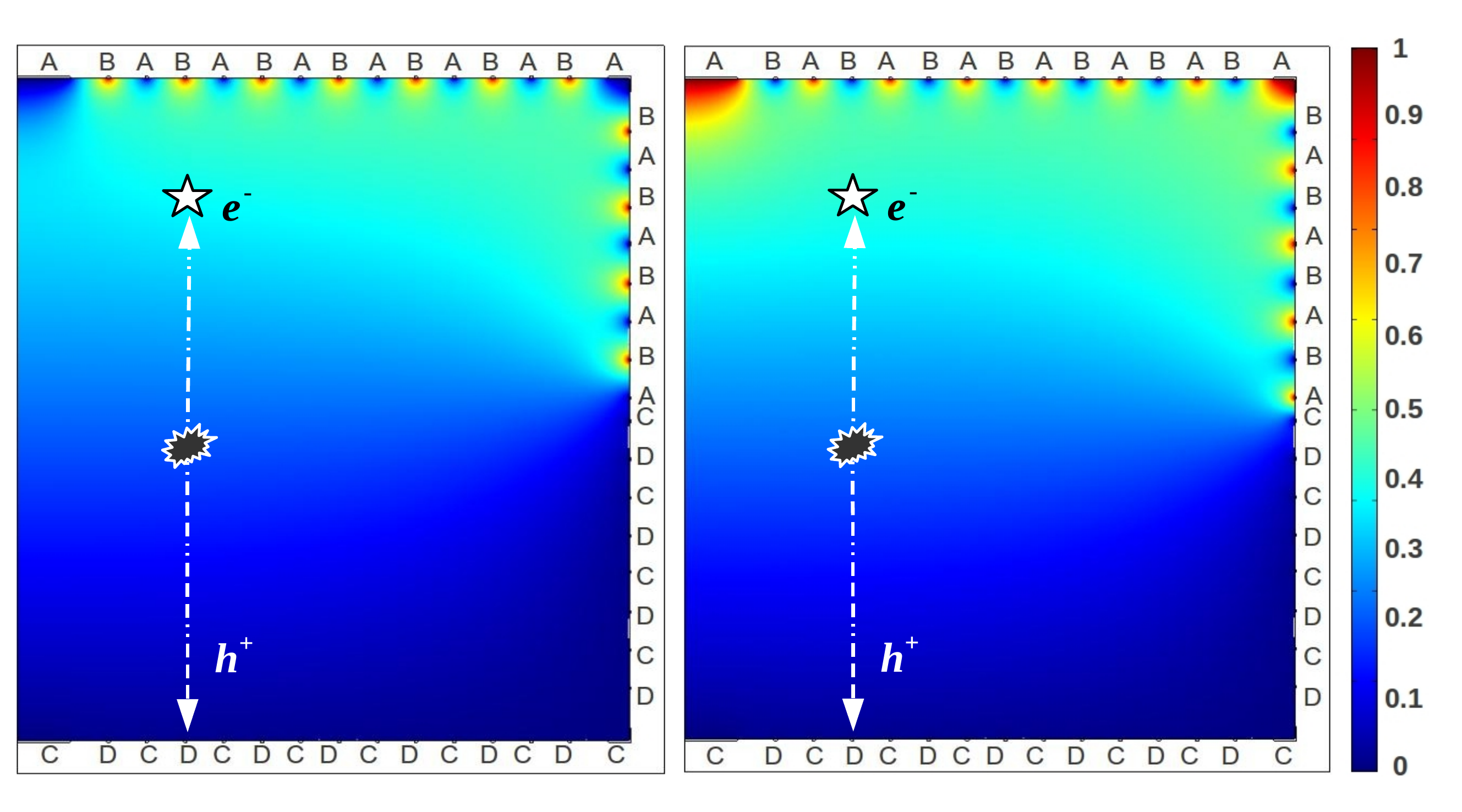}
\caption{\textit{Weighting potentials associated to electrodes B (left) and A (right) as indicated by the color code. 
The white star indicates the trapping location of the electron $\textbf{r}_{e\,\mathrm{F}}$ in the cyan region 
where $\phi_A(\textbf{r}_{e\,\mathrm{F}})=\phi_B(\textbf{r}_{e\,\mathrm{F}})=0.4$. 
The values $\phi_C(\textbf{r}_{e\,\mathrm{F}})=\phi_D(\textbf{r}_{e\,\mathrm{F}})=0.1$ can be deduced from weighting potentials associated to electrodes A and B in the dark blue region, 
at the location corresponding to the mirror image of $\textbf{r}_{e\,\mathrm{F}}$ from the equatorial axis.}}
\label{fig:RamoAC}
\end{figure}

The analytical model is based on the Shockley-Ramo theorem \cite{ramoinitial,ramo1939,ramobonnedemo}. 
We refer to \cite{ramoreview} for a comprehensive review of its various applications. This theorem states that the charge $Q_K$ induced on an electrode $K$ by a moving charge $q$ from a position $\textbf{r}_{q\,\mathrm{I}}$ to  $\textbf{r}_{q\,\mathrm{F}}$ is given by the relation:
\begin{equation}
Q_K=-q\left(\phi_K(\textbf{r}_{q\,\mathrm{F}})-\phi_K(\textbf{r}_{q\,\mathrm{I}} )\right)
\end{equation}
where $\phi_K(\textbf{r})$ is a dimensionless weighting potential obtained by setting the electrode $K$ to 1 and all the other electrodes to 0. In semi-conductors, $\phi_K(\textbf{r}_{e\,\mathrm{I}})=\phi_K(\textbf{r}_{h\,\mathrm{I}})$ as electrons and holes are created by pairs following an energy deposit. Therefore, the induced charge only depends on carrier positions at the end of their drift:
\begin{equation}
Q_K=e\left(\phi_K(\textbf{r}_{e\,\mathrm{F}})-\phi_K(\textbf{r}_{h\,\mathrm{F}} )\right)
\label{eqramointerest}
\end{equation}
Let's consider the simple application example of Eq.~(\ref{eqramointerest}) as schematised in Fig.~\ref{fig:RamoAC}: 
a fiducial event with a single $e^-/h^+$ pair created where the electron is trapped during its drift at the location indicated by the white star. 
Weighting potential maps associated to electrodes A and B are shown in Fig.~\ref{fig:RamoAC}. 
Those associated to electrodes C and D can be deduced from the equatorial symmetry. 
In this example, we obtain from of Eq.~(\ref{eqramointerest}):
\begin{eqnarray}
Q_A=e\left(\phi_A(\textbf{r}_{e\,\mathrm{F}})-\phi_A(\textbf{r}_{h\,\mathrm{F}})\right)=e\left(0.4-0 \right)=+0.4 e \nonumber \\ 
Q_B=e\left(\phi_B(\textbf{r}_{e\,\mathrm{F}})-\phi_B(\textbf{r}_{h\,\mathrm{F}})\right)=e\left(0.4-0 \right)=+0.4 e \nonumber \\ 
Q_C=e\left(\phi_C(\textbf{r}_{e\,\mathrm{F}})-\phi_C(\textbf{r}_{h\,\mathrm{F}})\right)=e\left(0.1-0 \right)=+0.1 e \nonumber \\
Q_D=e\left(\phi_D(\textbf{r}_{e\,\mathrm{F}})-\phi_D(\textbf{r}_{h\,\mathrm{F}})\right)=e\left(0.1-1 \right)=-0.9 e
\end{eqnarray}
This simple example illustrates that veto signals are expected even for fiducial events in the case of charge carrier trapping in the bulk. 
Also, the trapped carriers induce equivalent signals on a fiducial electrode and its neighboring veto electrode due to the similarity of the corresponding weighting fields. 
 \newline
Finally, one can notice that the total induced charge $Q_{tot}=Q_A+Q_B+Q_C+Q_D=0$ in the given example. 
This would be a natural result in the absence of trapping as both charges of opposite sign would be collected. 
However, the consequence of carriers trapped in the bulk during their drift is less obvious. 
Still, we can show that the sum of the signals induced by $e^-/h^+$ pairs is always zero. 
To demonstrate this charge conservation relation for trapping-induced signals, 
let's calculate the total induced charge $Q_{tot}$ using the Shockley-Ramo theorem. 
As $Q_K$ depends linearly on $\phi_K$, $Q_{tot}$ will only depend on the weighting potential 
$\phi_{tot}=\phi_A+\phi_B+\phi_C+\phi_D$, obtained by  setting all four sets of electrodes to 1.
Due to the germanium high permittivity ($\epsilon_r=16$) and the small spacing (2 mm) between electrodes, 
most of the field lines are not able to escape the system (absorber + electrodes). 
The resulting weighting potential $\phi_{tot}$ is very close to one in all the absorber. 
Applying the Shockley-Ramo theorem:
\begin{equation}
Q_{tot}=\sum_{n=1}^{N_p} e \left(\phi_{tot}(\textbf{r}_{e\,\mathrm{F}})_n-   \phi_{tot}(\textbf{r}_{h\,\mathrm{F}})_n\right)
= eN_p (1-1)=0
\label{conservation}
\end{equation}
where the label $n$ refers to the considered $e^-/h^+$ pair among the $N_p$ created. 
As equation (\ref{conservation}) is always verified, it implies that among the four ionization energy measurements, only three are independent. We will show in section  \ref{sec:appchargeconserv} how this property is now used by EDELWEISS to improve individual ionization baseline resolutions.
\begin{figure}[h!] 
\centering 
\includegraphics[width=0.45\textwidth,keepaspectratio]{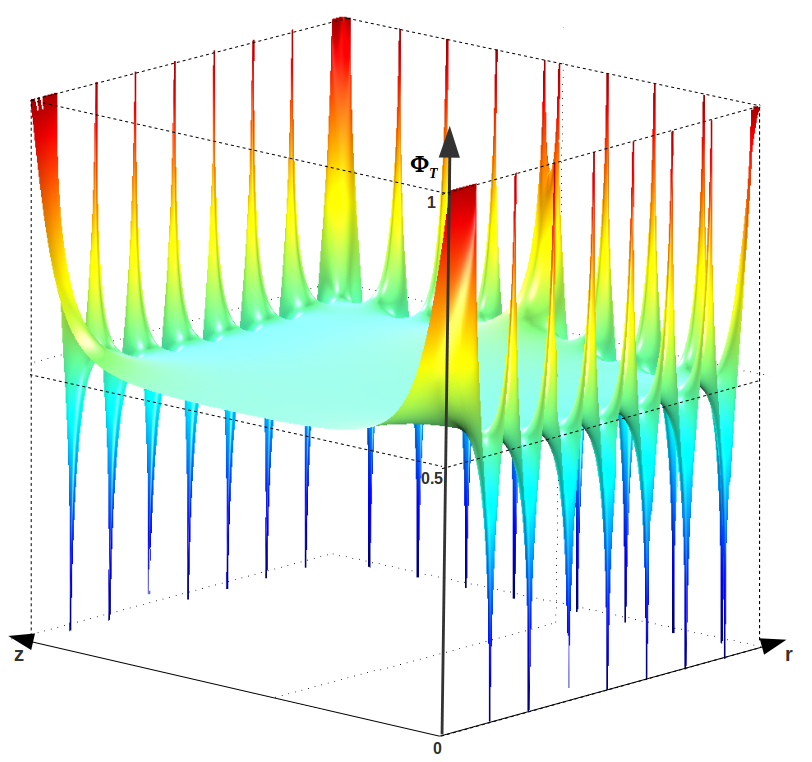}
\caption{\textit{3D representation of the weighting potential map $\phi_{A+C}$.}}
 \label{fig:weightingac}
 \end{figure} \newline
Another interesting property of the weighting fields results from the electrode geometry configuration. 
The sum of the weighting potentials associated to veto electrodes is constant in almost all the detector volume ($\phi_A+\phi_C\sim 0.5$) as shown on Fig.~\ref{fig:weightingac}.
Neglecting trapping in the remaining near-surface area, the sum of the signals induced on veto electrodes A and C for a fiducial event is given by:
\begin{equation}
Q_A+Q_C=\sum_{n=1}^{N_p} e \left(\phi_{(A+C)}(\textbf{r}_{e\,\mathrm{F}})_n- \phi_{(A+C)}(\textbf{r}_{h\,\mathrm{F}})_n\right) 
= 0.5 e (N_{\mathrm{T_e}}-N_{\mathrm{T_h}})
\label{eq05}
\end{equation}  
where $N_{Te}$ and $N_{Th}$ are the number of electrons and holes trapped during their drift.
We can see that $Q_A+Q_C$ linearly depends on the difference of the number of trapped carriers of each type. As a result, these signals must be highly correlated to both trapping lengths and energy deposit location, a statement that motivates an analytical modeling of this phenomenon.

\section{Analytical model}

\begin{figure}[t] 
\centering 
\includegraphics[width=0.55\textwidth,keepaspectratio]{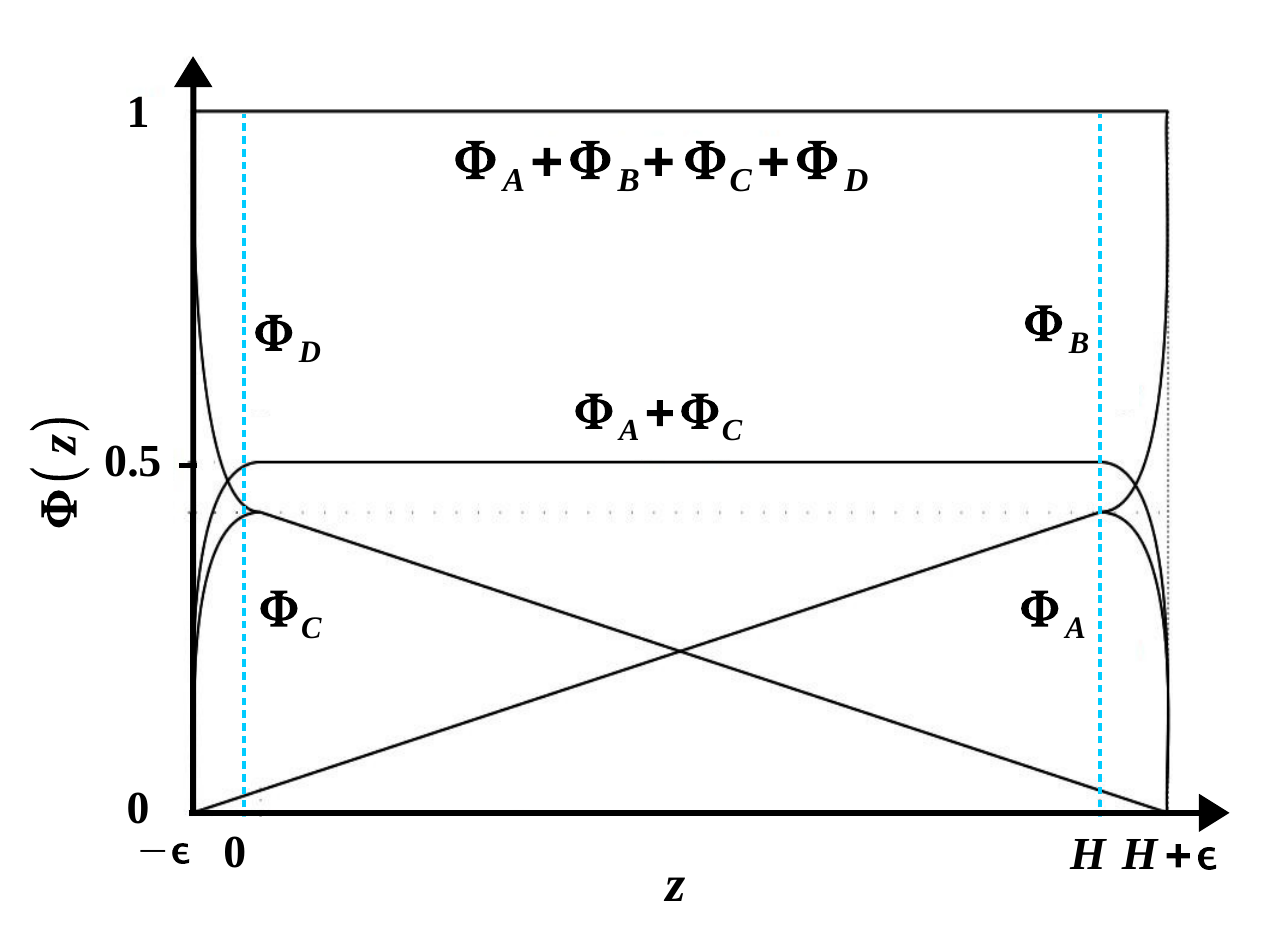}
\caption{\textit{scheme of relevant weighting potentials along a $z$-oriented axis linking two fiducial electrodes B and D.}}
 \label{fig:schemasmodeles}
 \end{figure}

The analytical model applies to fiducial events occurring in the near-homogeneous electric field region 
where it will be approximated to be constant. 
This further justifies the use of fixed mean trapping lengths for electrons $(l_e)$ and holes $(l_h)$ in their drift along a $z$-axis linking two fiducial electrodes B and D. 
Fig.~\ref{fig:schemasmodeles} represents the weighing potentials along this axis, where the two detector surfaces are
located at the coordinates $-\epsilon$ and $H+\epsilon$, and the field is constant in the interval $z\in[0,H]$.
The $N_p$ created $e^-/h^+$ pairs following an energy deposit at a depth $z=Z_0$ 
will drift to their corresponding fiducial collecting electrodes.
The number of remaining untrapped charge carriers of each type $j=\{e,h\}$ after a travelled distance $d=|z-Z_0|$ 
is thus given by $N_j=N_pe^{-\mu_jd}$ where $\mu_j=1/l_ j$ is the inverse of the trapping length. 
We assume that once charge carriers reach the near surface region ($z\notin[\mathrm{0},H]$), 
they can be considered as collected for three reasons. First, due to a grid effect, the surface electric field is an order of magnitude more intense than in the bulk which means that trapping should be comparatively small in most of the non-fiducial volume. Second, weighting potentials approaching the fiducial electrodes rapidly reach 0 for veto electrodes and 1 for fiducial electrodes as shown in Fig.~\ref{fig:schemasmodeles}. Finally, the remaining drift distance $\epsilon\sim 0.2\;\mathrm{cm}$ is small relative to the bulk length $H\sim 3.6\;\mathrm{cm}$. 
We will thus consider only the contribution of bulk-trapped carriers to the total charge induced on an electrode $K=\{A,B,C,D\}$, obtained through the application of the Shockley-Ramo theorem:
\begin{equation}
Q_K=e\int_{Z_0}^{H}\mu_eN_pe^{-\mu_e(z-Z_0)}\phi_K(z)\mathrm{dz} 
 -  e \int_{0}^{Z_0}\mu_hN_pe^{\mu_h(z-Z_0)}\phi_K(z)\mathrm{dz} 
\label{generalformula}
\end{equation}
As $\phi_K(z)$ is obtained by setting to 0 all the electrodes of one side and alternatively to 0 and 1 the ones of the other side (0.5 in average), 
the weighting potential rises with a slope of $0.5/L$ (where $L=H+2\epsilon=4\;\mathrm{cm}$ is the detector length) in the region $z\in[0,H]$ where the analytical model applies. 
Also, the weighting potential associated to a veto electrode and its neighbouring fiducial electrode are the same such that:
\begin{equation}
\phi_{(A,B)}=\frac{z+\epsilon}{2L} \quad \quad \phi_{(C,D)}=\frac{-z+H+\epsilon}{2L}
\label{weightingindividual}
\end{equation} 
By substituting Eq.~(\ref{weightingindividual}) in Eq.~(\ref{generalformula}), 
we get the charge induced on any electrode by bulk-trapped charges as a function of the trapping lengths and the deposit depth $Z_0$.

\subsection{Veto signals induced by trapped charges}

\label{vetosignals}
For fiducial events, no net charge is collected on the veto electrodes A and C, 
and thus Eq.~(\ref{generalformula}) corresponds to the total expected ionization signal. 
Exact trapping lengths are unknown and may vary from one detector to another depending on the purity of the crystals \cite{MCPiroLTD}. 
For normal biasing conditions, they are expected to be large in comparison with the bulk length $H$, 
allowing a series expansion in $\mu_j H \ll 1$. 
Interesting properties appear when considering the following first-order expansion:
\begin{equation}
Q_A+Q_C=-\frac{eN_p}{2}\left(Z_0(\mu_e+\mu_h)-\mu_e H \right)
\label{eqn-plus}
\end{equation}
\begin{equation}
Q_A-Q_C=-\frac{eN_p}{2L}(\mu_e+\mu_h)\left(Z_0^2-Z_0H \right)
\label{eqn-minus}
\end{equation}
The sum of the induced charges on veto electrodes linearly depends on $Z_0$ due to the weighting field property previously underlined through Eq.~(\ref{eq05}), whereas a quadratic dependence appears when we subtract them. In other words, these veto electrodes that didn't collect a single charge may be used to provide an estimate of the energy deposit depth. Also, fiducial events should be distributed along a parabola in the plane $(Q_A+Q_C,Q_A-Q_C)$.   

\subsection{Evaluation of the model}

To evaluate the relevance of analytical model assumptions, we first confront it to a numerical model. 
In the latter, both electric and weighting fields are computed using a finite element evaluation of the Poisson equation. 
Single mono-energetic deposits are homogeneously distributed in the detector, 
and fiducial events are defined by the absence of charge carrier reaching the veto electrodes, 
instead of an arbitrary depth as in the analytical model. 
Charge carriers drift along the actual electric field lines, and not simply along the $z$-axis 
(although transport anisotropies \cite{AlexBLTD} are not taken into account).
The trapping lengths vary according to the field intensity \cite{MCPiroLTD}, 
and in particular, the trapping of charges close to the surface is not neglected. 

\begin{figure}[h!] 
\centering 
\includegraphics[width=0.44\textwidth,keepaspectratio]{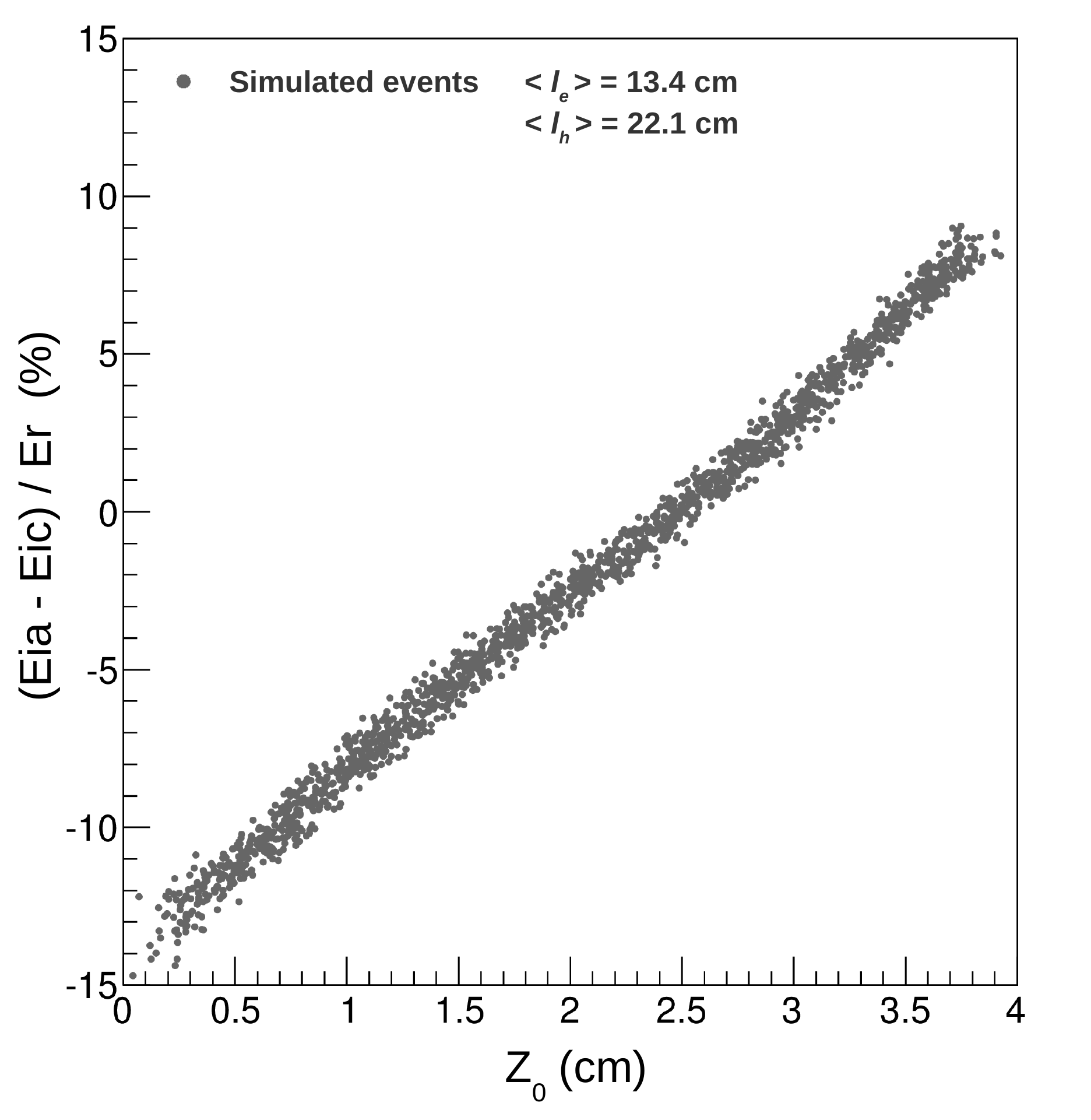}
\includegraphics[width=0.44\textwidth,keepaspectratio]{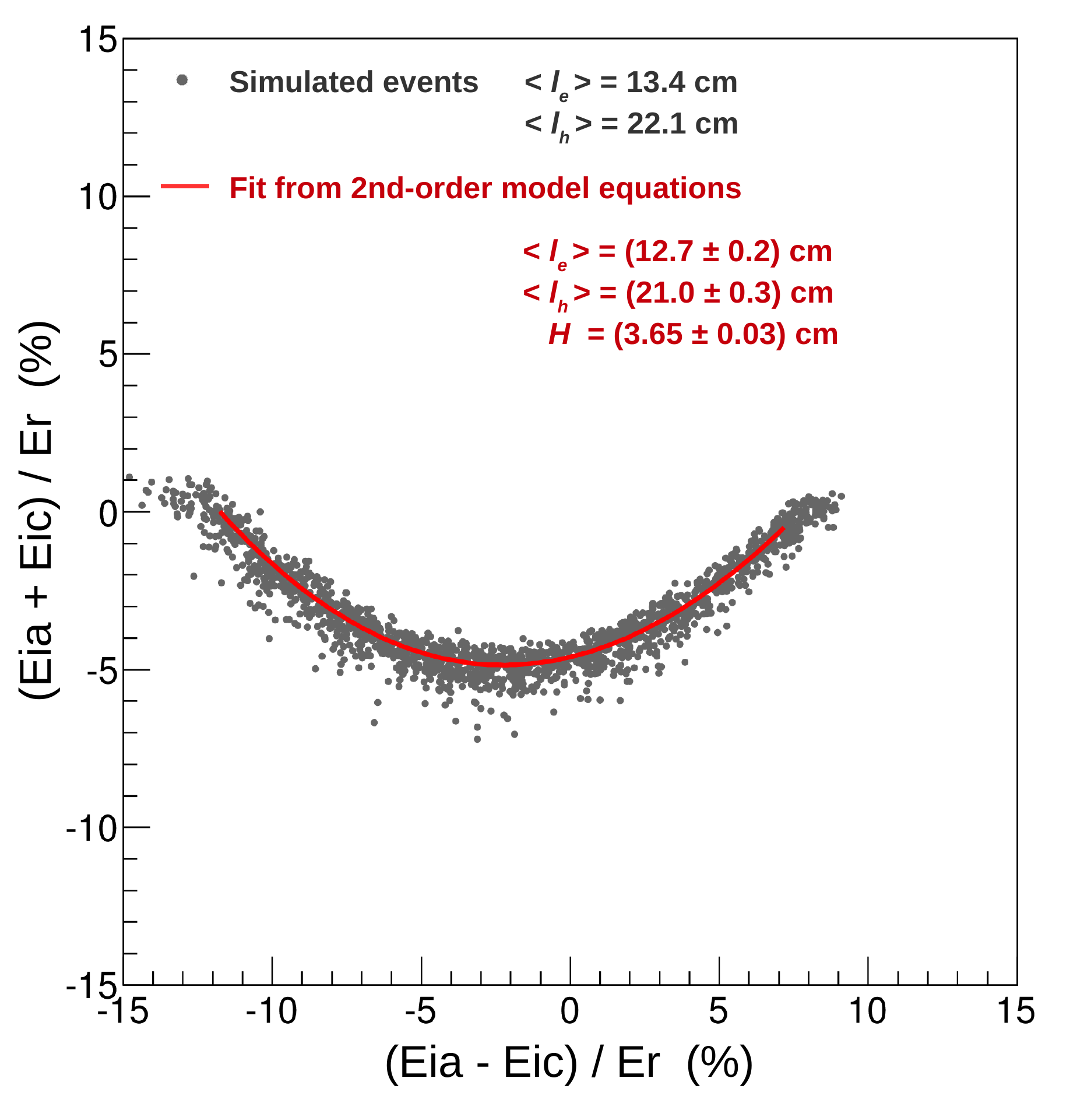}
\caption{\textit{Fiducial events in the region where the field is nearly homogeneous (radii between 0.5 and 2.5 cm)
 simulated with the numerical model. 
Top: check of the linear dependence of $(Eia-Eic)/Er$ on the deposit depth $Z_0$. 
Bottom: Distribution of $(Eia+Eic)/Er$ as a function of $(Eia-Eic)/Er$ for these events, together with a
fit  (red line) of the second-order expansion of the model.
Values implemented in the numerical model are shown in black and the ones obtained from the fit are shown in red.   }}
 \label{fig:depthandparabolla}
 \end{figure}

The output of the numerical simulation are the charge signals $Q_K$ on the four electrodes for
a given number of initial $e^-/h^+$ pairs $N_p$.
The calibrated experimental data for  an event with a recoil energy $Er$
consist of the four ionization energy signals $E_{iK}$ (defined as
positive in absence of trapping), according to the expressions:
\begin{equation}
\frac{Eia}{Er}=-\frac{Q_A}{eN_p} \quad \quad \quad  \frac{Eib}{Er}=\frac{Q_B}{eN_p} \quad \quad \quad  \frac{Eid}{Er}=-\frac{Q_D}{eN_p}  \quad  \quad \quad  \frac{Eic}{Er}=\frac{Q_C}{eN_p}
 \label{expressions}
\end{equation}
Figure \ref{fig:depthandparabolla} (top) shows the distribution of $(Eia-Eic)/Er$ as a function
of $Z_0$ for simulated fiducial events located at radii between 0.5 and 2.5 cm.
One can see that the linear dependence on $Z_0$ predicted by the first-order expansion (Eq.~(\ref{eqn-plus})) 
is well confirmed by the numerical simulation.
The simulated signals are not smeared by some experimental resolution:
the observed spread and small non-linearities on this figure are due to the varying length of the curved field lines,
and also the curvature of the boundary between the fiducial and surface regions.
This figure shows that, given a sufficient ionization energy resolution, the quantity
$(Eia-Eic)/Er$ can be used to measure the depth of localized energy deposits inside the region
of the fiducial volume where the field is nearly uniform\footnote{
This simple behavior gets distorted when selecting bulk events in the region where the
field is highly inhomogeneous (as along the equator, see Fig.~\ref{fig1}),
and provisions for such effects will be taken when the model will be compared with data.}.
Figure \ref{fig:depthandparabolla} (bottom) shows the distribution of the same events in the 
plane $((Eia+Eic)/Er,(Eia-Eic)/Er)$. 
According to the first-order expansion (Eqs.~(\ref{eqn-plus}) and~(\ref{eqn-minus})),
the events should be distributed along a parabola.
Here again, the numerical simulation confirms such behavior.
The agreement between the analytical and numerical models can even be compared
quantitatively, provided that the expansion is extended to the second order in $\mu_jH$.
The red line in Fig.~\ref{fig:depthandparabolla} (bottom) represents the fit of the distribution of simulated events 
to the second-order development of Eqs.~(\ref{generalformula}) and~(\ref{weightingindividual}),
with the trapping lengths $l_e=1/\mu_e$, $l_h=1/\mu_h$ and the bulk length $H$ as free parameters. 
We retrieve the input trapping lengths associated to the average field intensity in the bulk 
($<{l_e(||\vec{E}||)}>=13.4\;\mathrm{cm}$ and $<l_h(||\vec{E}||)>=22.1\;\mathrm{cm}$ where $||\vec{E}||=0.625\;\mathrm{V/cm}$) 
within a precision of 5\%: 
$l_e=(12.7\pm 0.2) \;\mathrm{cm}$,  
$l_h=(21.0\pm 0.3)\;\mathrm{cm}$ and $H=(3.65\pm0.03)\;\mathrm{cm}$ where errors are statistical only. 
This not only denotes that the assumptions of the analytical model are justified 
but also shows that some trapping information may be extracted from data. 

For the ease of the comparison of the data with the model (see Sec.~\ref{section-appli}), 
it is preferable to fit $((Eia+Eic)/Er,(Eia-Eic)/Er)$ distributions with the simple parabolic dependence
predicted by the first-order expansion of Eqs.~(\ref{eqn-plus}) and~(\ref{eqn-minus}):
%\begin{eqnarray}
%\frac{Eia+Eic}{Er}&=&\frac{1}{L(\mu_e+\mu_h)}\left[2\left(\frac{Eia-Eic}{Er}\right)^2
% \right. \nonumber \\ 
%&+&\left. H(\mu_e-\mu_h)\left(\frac{Eia-Eic}{Er}\right) \right.
%-\left. \mu_e\mu_h\frac{H^2}{2}\right]
%\label{eqapcamc}
%\end{eqnarray}
\begin{eqnarray}
\frac{Eia+Eic}{Er}&=&\frac{2}{L(\mu_e+\mu_h)}\left[\left(\frac{Eia-Eic}{Er}\right)^2+\frac{H}{2}(\mu_e-\mu_h)\left(\frac{Eia-Eic}{Er}\right)- \mu_e\mu_h\frac{H^2}{4}\right]
\label{eqapcamc}
\end{eqnarray}

A fit of  that expression to the distribution in Fig.~\ref{fig:depthandparabolla} (bottom), 
with $H$ fixed to the value of 3.65 cm,
recovers the true values $\left<l_e\right>$ and $\left<l_h\right>$ to within 20\%. 
The first-order equation~(\ref{eqapcamc}) is thus sufficient to account for the main trapping effects.

\subsection{Ionization signals on fiducial electrodes}

\label{subsec:ionsignals}
The expected signals on the fiducial electrodes B and D come from two contributions: charge collected at the electrodes and induced charge by trapped carriers. The latter is given by Eq.~(\ref{generalformula}) as trapped charges induce equivalent signals on a veto electrode and its neighbouring fiducial electrode. For its part, the collected charge consists of untrapped carriers reaching the near-surface region. Thus, the total ionization signals on fiducial electrodes B and D are given by:
\begin{eqnarray}
Q_B&=&Q_A+eN_pe^{-\mu_e(H-Z_0)} \label{qb}\\
Q_D&=&Q_C-eN_pe^{-\mu_hZ_0} \label{qd}
\end{eqnarray} 

\begin{figure}[t] 
\centering
\includegraphics[width=0.44\textwidth,keepaspectratio]{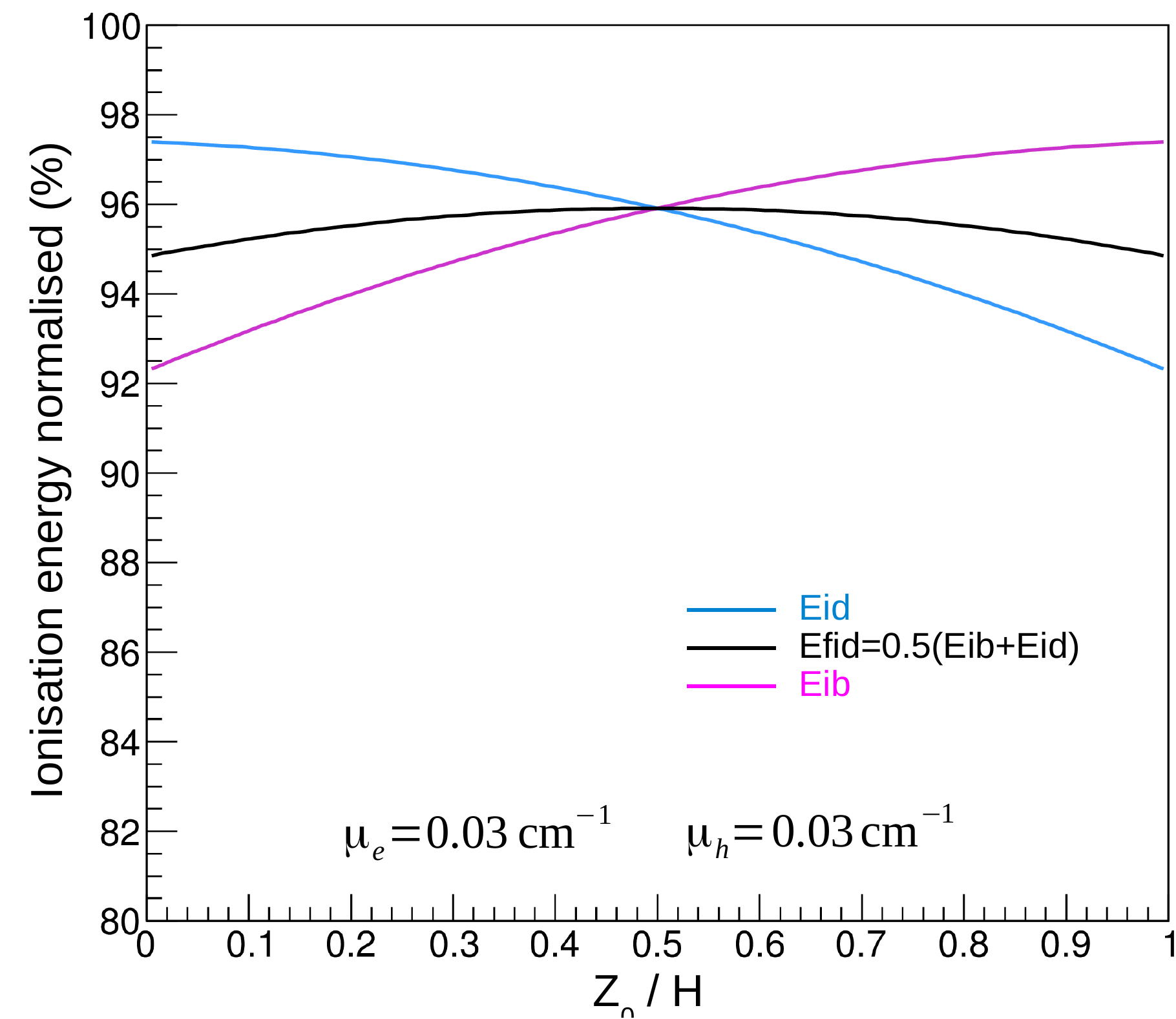}
\includegraphics[width=0.44\textwidth,keepaspectratio]{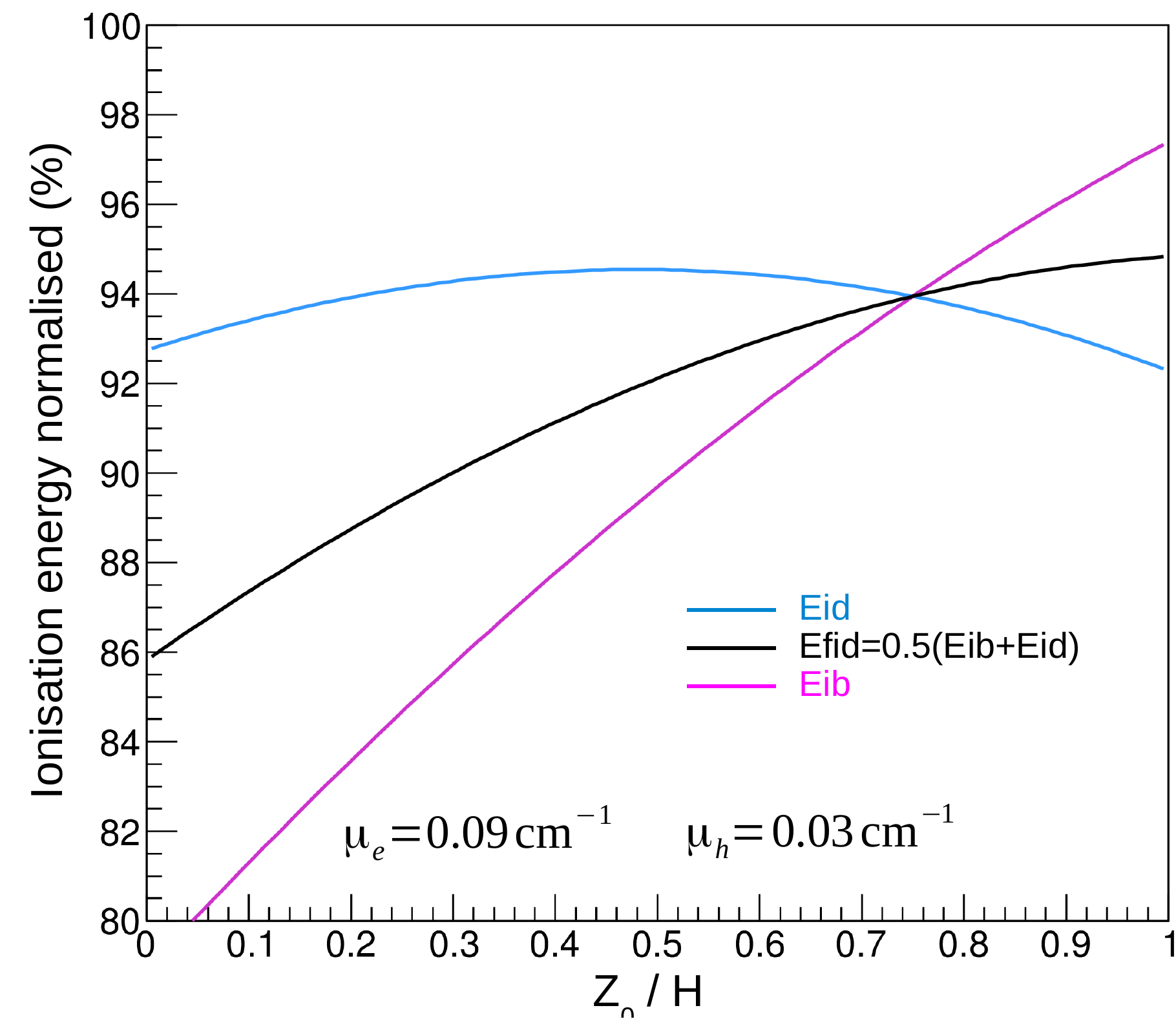}
\caption{\textit{Trapping impact on ionization measurements according to the model equations. Ionization energies are normalized to unity for a complete charge collection and shown as a function of the deposit depth $Z_0$ normalized to the bulk length $H$ for the cathode (blue), the anode (pink) and the fiducial ionization energy (black). Trapping lengths considered are shown at the bottom of each panel.}}
 \label{fig:qbqd}
 \end{figure}

Using equations (\ref{expressions}), (\ref{qb}) and (\ref{qd}), 
we represent on Fig.~\ref{fig:qbqd} the ionization energies, normalized to unity in case of complete charge collection, as a function of the deposit depth for two different scenarios: 
in Fig.~\ref{fig:qbqd} (top) electrons and holes are similarly trapped ($\mu_e=\mu_h=0.03\;  \mathrm{cm^{-1}}$)  
whereas in Fig.~\ref{fig:qbqd} (bottom), 
electron trapping is significantly increased ($\mu_e=3\mu_h=0.09\;  \mathrm{cm^{-1}}$). 
Blue and pink lines correspond to the normalized signals of the fiducial cathode ($Eid/Er$) and anode ($Eib/Er$). 
The black line represents ($Efid/Er$) where $Efid$ is the fiducial ionization energy defined as $Efid=0.5(Eib+Eid)$.
One can appreciate the flat shape of $Efid/Er$ for $\mu_e=\mu_h$ that clearly leads to only tiny energy dispersions $<1 \%$ whereas those expected from $Eib/Er$ or $Eid/Er$ are $>5 \%$. 
Now, when the electron trapping is increased by a factor three, we see that the resolution associated to $Eib/Er$ and $Efid/Er$ is significantly degraded, which is typically the effect one could expect. 
However, it is very interesting to see that in this special case, 
the worse electron trapping improves the resolution of the fiducial cathode D that collects holes. 
This effect can be analytically explained by assessing the width of the energy distribution through the difference between the expected energies for the extremum values of the deposit depth ($Z_0=0$ and $Z_0=H$). 
In the first-order expansion, $\Delta_{Eid}=Eid(0)-Eid(H)$ is simply proportionnal to $(3\mu_h-\mu_e)$, 
while the corresponding quantity for the anode, $\Delta_{Eib}$, is proportional to $(3\mu_e-\mu_h)$. 
Thus, $\Delta_{Eid}$ vanishes in the second scenario. 
Similarly we are able to justify the small fiducial energy dispersions observed in the first scenario 
as $\Delta_{Efid}\propto (\mu_e-\mu_h)$.

This study clearly shows that the choice of the best ionization-based energy estimator must be taken with a careful understanding of trapping effects as both electrons and holes affect the signals read on B and D channels. 
Though we identify the individual electrode that provides the best resolution as the one collecting the least trapped of the two carriers, 
an improvement is expected when using a weighted sum of B and D signals, 
where weights may vary from one detector to an other depending on trapping properties of the crystal. 
As we will see in section \ref{sec:corrections}, 
an even more precise measurement can be achieved by cancelling out the deposit depth dependence with a correction function based on the veto signals.

\subsection{Trapping impact on heat signal}

We now model the impact of charge-trapping on the heat signal.
For $N_p$ created $e^-/h^+$ pairs, the Neganov-Luke heating is given by the following relation:
\begin{equation}
E_{Luke}=e\sum_{n=1}^{N_p}\left( V(\vec{r_e})_n-V(\vec{r_h})_n\right)
\end{equation}
where $V(\vec{r_e})_n$ and $V(\vec{r_h})_n$ are respectively the electric potentials at the end of the drifts of the electron and hole of the considered $n^{th}$  pair. The contribution of trapped charges to this heating $E^T_{Luke}$ is obtained similarly as in Eq.~(\ref{generalformula}) by replacing the weighting potential $\phi_K(z)$ with the electric potential $V(z)$:
\begin{equation}
E^T_{Luke}=e\int_{Z_0}^{H}\mu_eN_pe^{-\mu_e(z-Z_0)}V(z)\mathrm{dz} 
 -  e \int_{0}^{Z_0}\mu_hN_pe^{\mu_h(z-Z_0)}V(z)\mathrm{dz} 
\label{formulaheattrapped}
\end{equation}
In FID detectors, electrodes are biased such that $V_B=-V_D$ and $V_A=-V_C$. 
The average value of electric potentials on the top surface is $(V_A+V_B)/2=V_M$ and $(V_C+V_D)/2=-V_M$ at the bottom. 
This leads to a homogeneous electric field $E_m=2V_M/L$ and therefore to a linear electric potential in the bulk region $z\in[\mathrm{O},H]$:
\begin{equation}
V(z)=\frac{V_M}{L}(2z-H)
\label{Vz}
\end{equation}
Substituting Eq.~(\ref{Vz}) in Eq.~(\ref{formulaheattrapped}), we get the contribution of bulk-trapped charges to Neganov-Luke heating. 
Since charge carriers that reach the near surface region can be considered as collected, 
the contribution of untrapped charges to the Neganov-Luke effect is given by:
\begin{equation}
E^C_{Luke}=N_p \left(V_B e^{-\mu_e(H-Z_0)}-V_De^{-\mu_hZ_0}\right)
\end{equation} 
The heat signal is then simply obtained by summing the different contributions:
\begin{equation}
E_{Heat}=E_R+E^T_{Luke}+E^C_{Luke}
\end{equation}
This analytical formulation can be used to assess the expected heat energy dispersion from trapping as we did in section \ref{subsec:ionsignals} for ionization. At first-order, $\Delta_{E_{Heat}}=E_{Heat}(0)-E_{Heat}(H)\propto (\mu_e-\mu_h)$. This can be considered unfortunate as electrons are expected to be systematically more trapped than holes at cryogenic temperatures \cite{MCPiroLTD}. Actually, the discrimination of electronic and nuclear recoils is based on the ratio of heat and ionization. It is thus the dispersion on the ratio heat/ionization that matters the most rather than the individual resolutions. 
\begin{figure}[t] 
\centering
\includegraphics[width=0.44\textwidth,keepaspectratio]{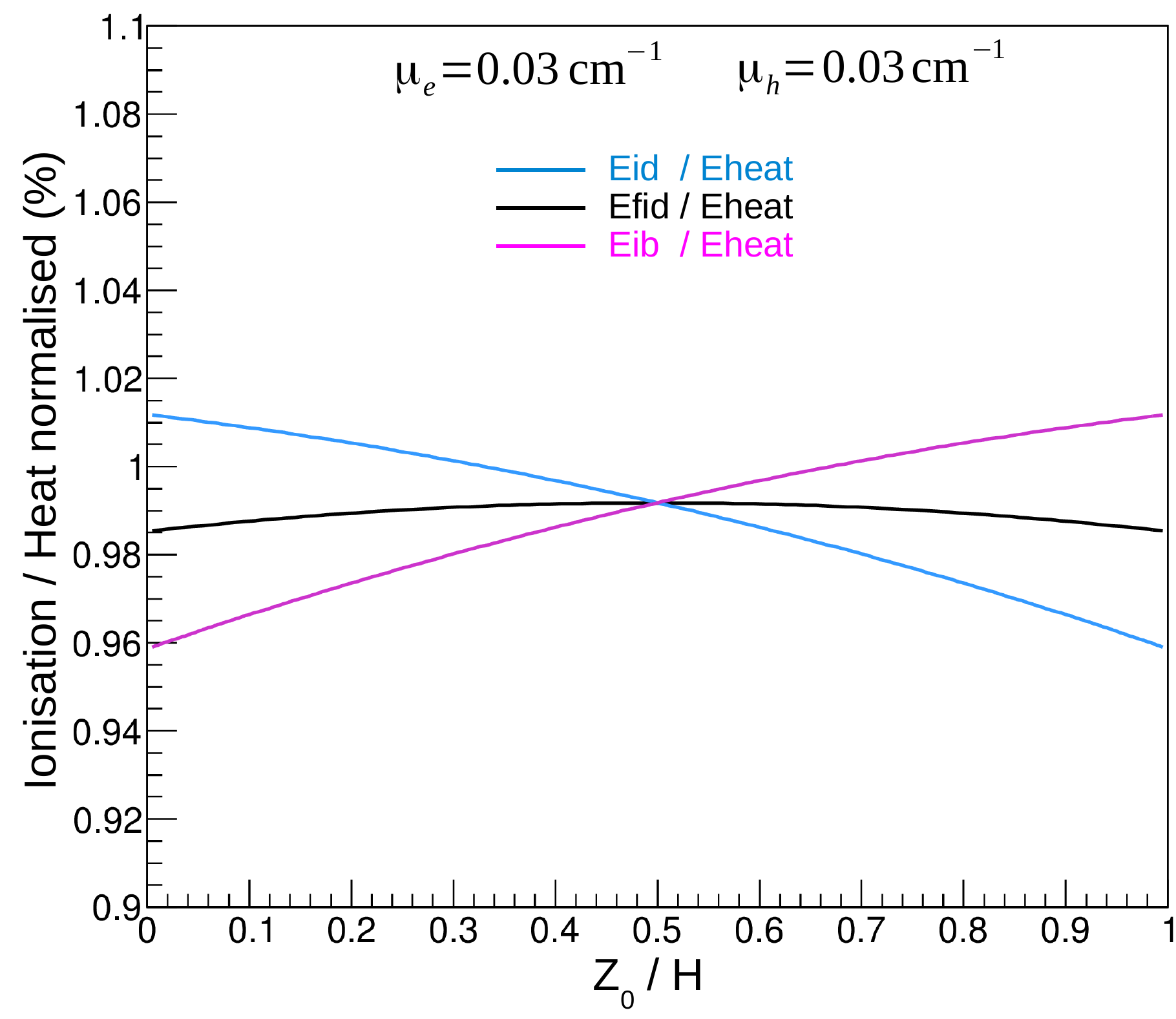}
\includegraphics[width=0.44\textwidth,keepaspectratio]{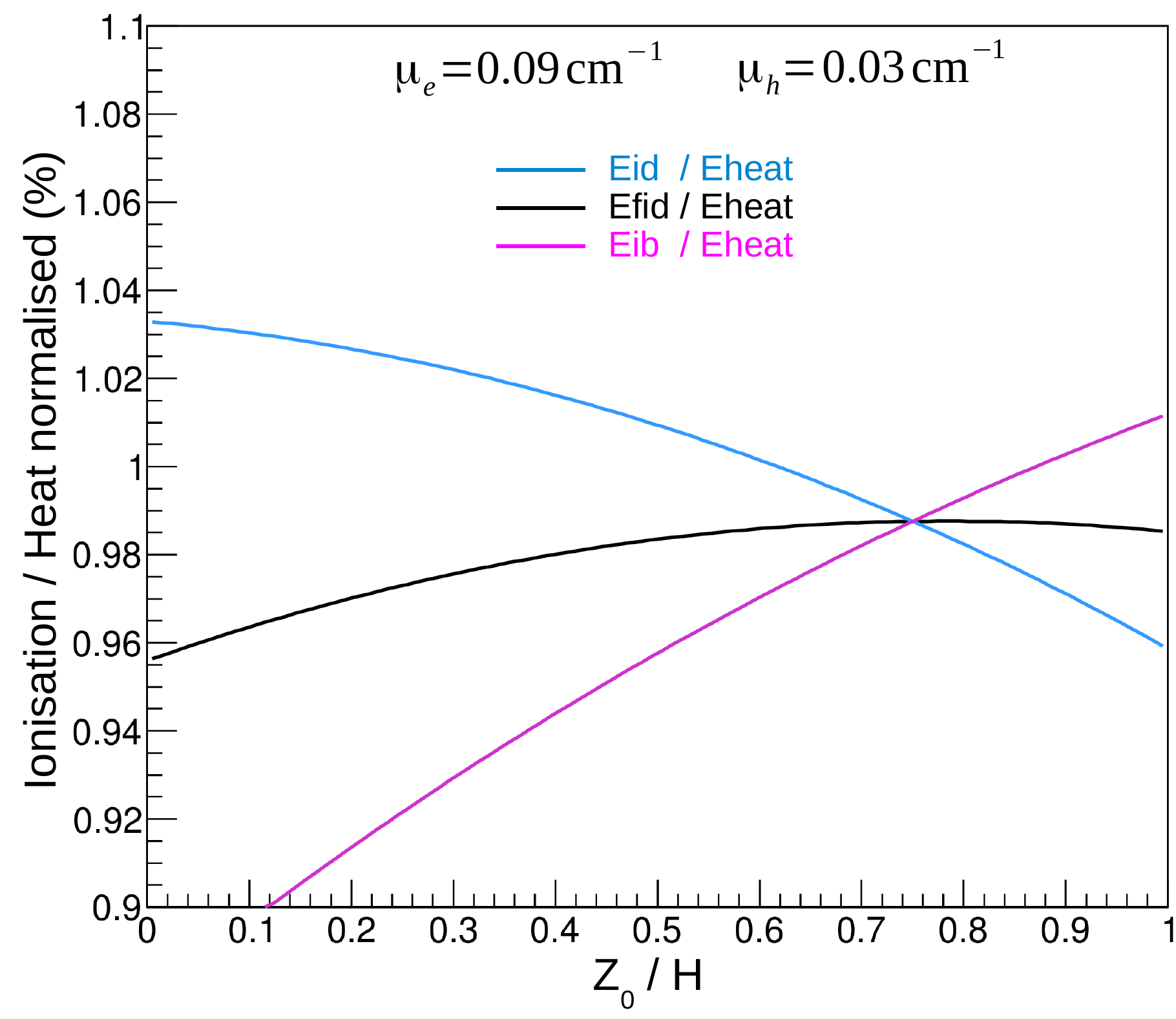}
\caption{\textit{same as Fig.~\ref{fig:qbqd} but considering the ratio ionization / heat normalized to unity for a complete charge collection.}}
 \label{fig:rapp}
 \end{figure}
Figure \ref{fig:rapp} shows the expected dependence of this ratio, normalized to unity in absence of trapping, 
for the same scenarios than earlier: $\mu_e=\mu_h$ (top panel) and $\mu_e=3\mu_h$ (bottom panel). 
Pink, blue and black lines respectively correspond to $Eib$, $Eid$ and $Efid$ used as estimators of the ionization energy. 
Clearly, the use of $Efid$ leads to much lower dispersions on the ratio in both cases due to the similar dependence in ($\mu_e-\mu_h)$ of $\Delta_{E_{Heat}}$ and $\Delta_{Efid}$. 
The ratio of $Efid$ and $E_{Heat}$ doesn't depend at first-order on the trapping parameters which attests of tight correlations between these measurements.

Thus, even if the best resolution on the absolute value of the ionization signal is provided by the channel with the smallest
trapping effects, we indentify $Efid$ as the best choice of ionization estimator when it is to be combined with the heat 
measurement, as done when calculating the relative ionization yield used in WIMP searches.

\section{Applications}
\label{section-appli}

\subsection{Characterization of bulk trapping in the crystals}

\begin{figure}[t] 
\centering 
\includegraphics[width=0.44\textwidth,keepaspectratio]{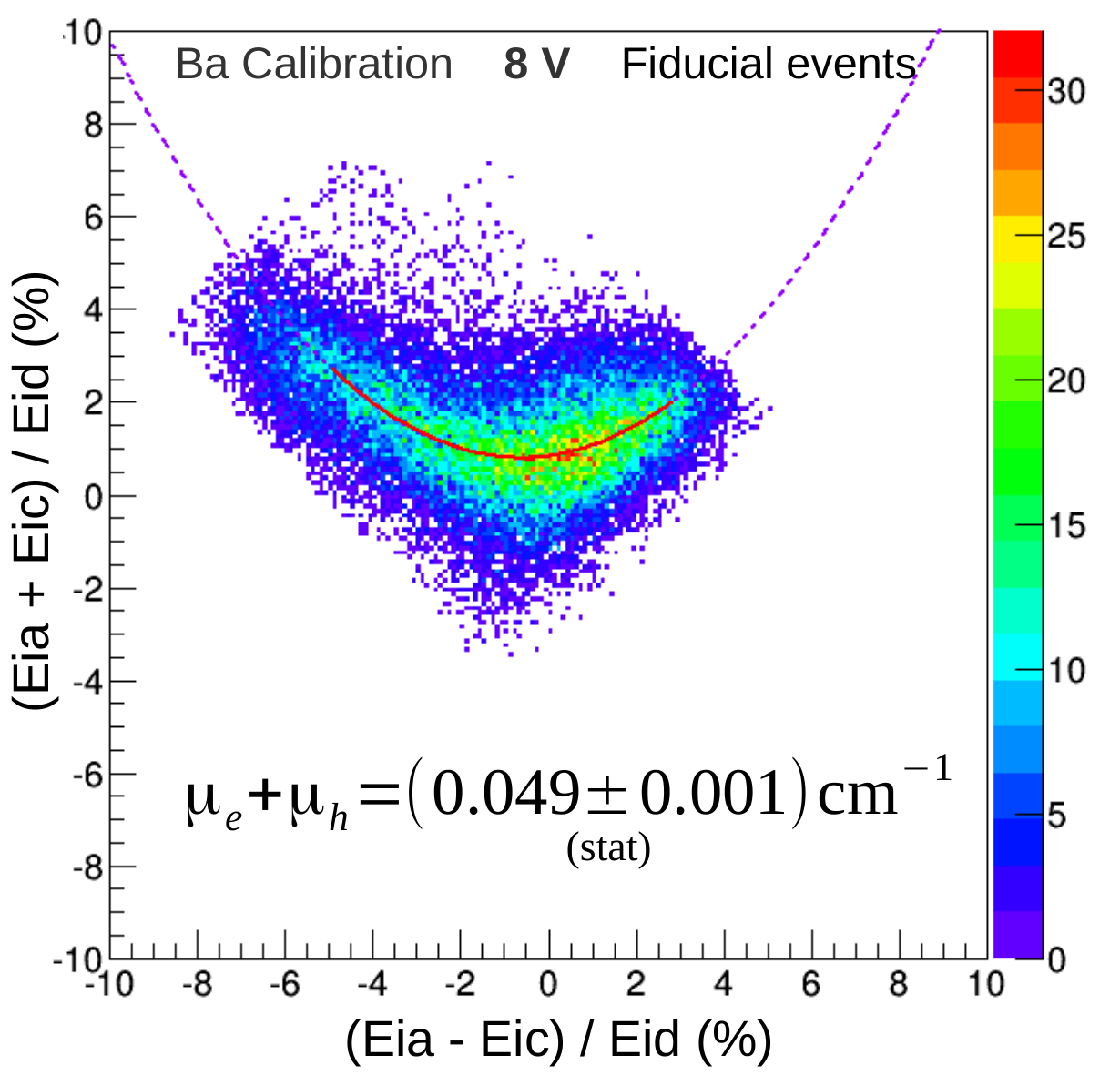}
\includegraphics[width=0.44\textwidth,keepaspectratio]{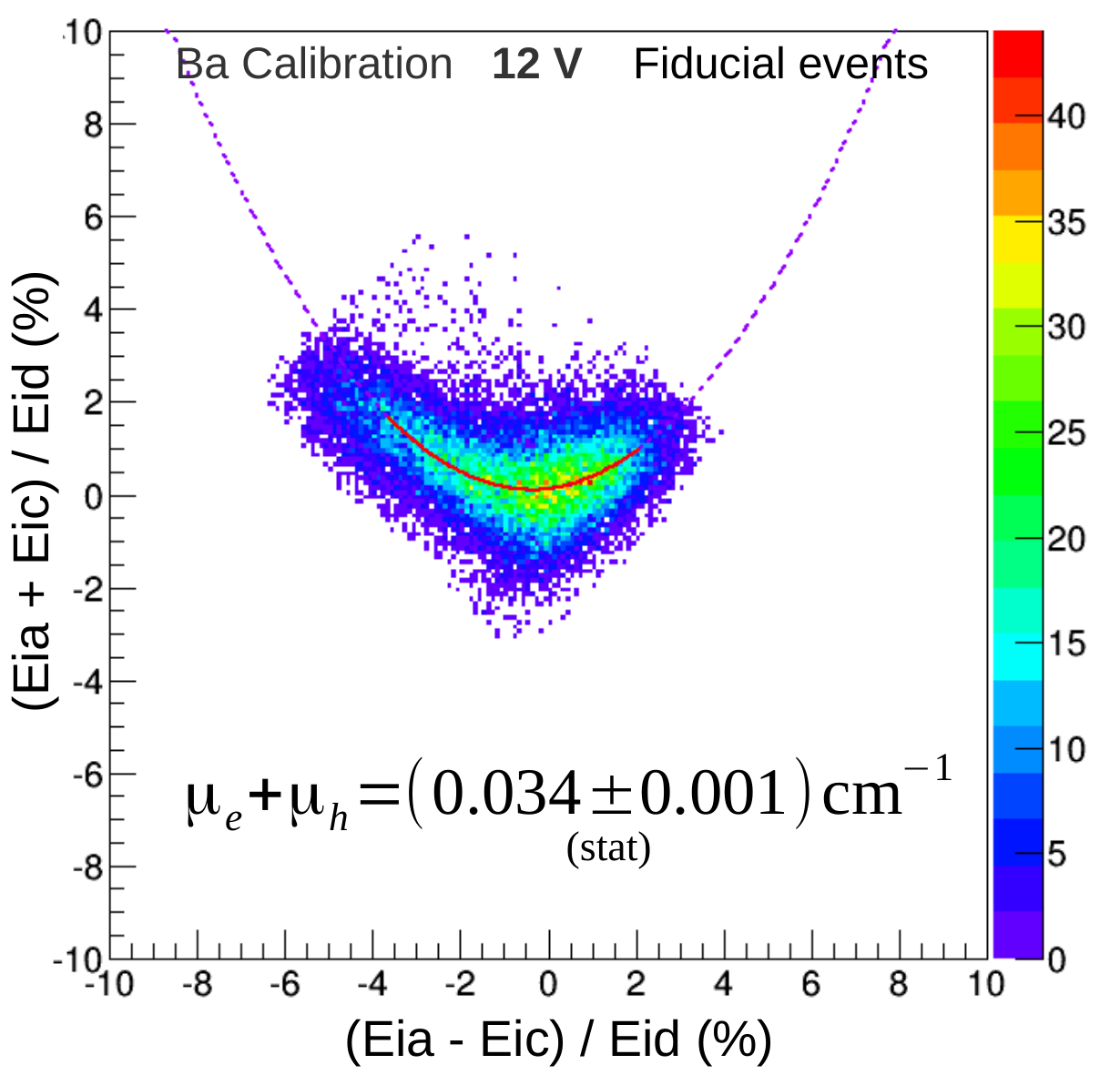}
\caption{\textit{Influence of field intensity on extracted trapping parameters. The top (bottom) panel shows the distribution of fiducial events ($>$100 keV) when operating at standard (50\% higher) bias voltage conditions. 
The dashed curve corresponds to a 2$^{nd}$-order polynomial fit to the distibution of events in the range  indicated by the solid red line. The value of the trapping indicator $\mu_e+\mu_h$ is derived from the fit using Eq.~(\ref{eqapcamc}).}}
 \label{fig:parabolledata}
 \end{figure}

We have seen that we were able to retrieve the trapping lengths implemented in the numerical simulation by fitting the distribution of simulated fiducial events in the plane of $(Eia+Eic)/Er$ as a function of $(Eia-Eic)/Er)$  with the analytical model equations. 
Unfortunately, in data, the parasitic capacitance between the interleaved electrodes, and between read-out channels,
induces cross-talk signals that can obscure the interpretation of the ratio of signals on different electrodes. 
During the calibration process, cross-talk correction coefficients are determined such that on average, 
veto signals are zero for fiducial events. 
In fiducial $\gamma$ events from $^{133}\mathrm{Ba}$ calibrations, as shown in Fig.~\ref{fig:parabolledata}, 
this results in distributions\footnote{Eid appears in the denominator here because it provides the estimator of the absolute energy $Er$ 
with the best resolution in cases where $\mu_e>\mu_h$, as encountered experimentally.}
where both $(Eia+Eic)/Eid$ and $(Eia-Eic)/Eid$ are also arbitrarily centered at approximately (0,0). 
As a result, the fit of Eq.~(\ref{eqapcamc}) to these data points results in somewhat arbitrary values for the constant and linear term.
The constant affecting the quadratic term, $\left((\mu_e+\mu_h)L/2\right)^{-1}$, is more robust.
Such a fit cannot disentangle the individual values of $\mu_e$ and $\mu_h$.
It is however still possible to extract important information on trapping, namely the sum of the inverse trapping lengths of
electron and holes, $\mu_e+\mu_h$.
Figure \ref{fig:parabolledata} shows the distribution of registered events for two different bias voltage conditions: 
the standard one (top) that results in an electric field in the bulk of $||\vec{E}||=0.625\;\mathrm{V/cm}$ and a $50\%$ higher (bottom) resulting in $||\vec{E}||=0.937\;\mathrm{V/cm}$. 
Note that only events above $100\;\mathrm{keV}$ are selected to make sure that the contribution of electronic noise is negligible. 
Trapping values $\mu_e+\mu_h$ are extracted from the fit of the second-order polynomial $y(Z_0)=a_0+a_1Z_0+a_2Z_0^2$,
with $a_2=2/L(\mu_e+\mu_h)$ and $Z_0=(Eia-Eic)/Eid$. 
As expected, we find that the trapping magnitude is reduced by increasing the electric field intensity. 
This effect has been checked on all detectors and is consistent with more direct measurements \cite{MCPiroLTD}. 
However, the extracted value from the fit should not be considered as a precise measurement of $\mu_e+\mu_h$ as the
fiducial region in the data also covers some part of the outer perimeter where both high- and low-field coexist. 
The simulations suggest that events from the high-field region near the equator are in part responsible for the
accumulation of counts in a point of the parabola that is not centered, due to trapping asymmetry, although the
non-uniform coverage the detectors by the gamma-ray flux from the calibration source plays also a role.
Nevertheless,  these effects in data do not affect strongly the dispersion of the events in $(Eia+Eic)/Eid$ relative to the 
parabola, and it was observed that consistent curvature results can be obtained using gamma rays of different origin
and energies (from 100 to 2000 keV).
The curvature is a good indicator of trapping in FIDs and is particularly useful to assess its importance even in
data sample where this effect cannot be studied by the position and width of visible peaks in the gamma-ray spectrum. 
It is now used in EDELWEISS as a quality criterium for the selection of detectors used for WIMP searches. \\

Finally, we clearly identify that, for fiducial events, 
the veto signals observed are primarily induced by charge carriers trapped in the bulk of the detector. 
The large energy-dependent dispersion of the veto ionization measurements pointed out in 
section~\ref{sec:intro} (cf. Fig.~\ref{fig:coupfid}) is thus attributable to charge trapping in the fiducial volume. 
Also, since events are distributed along a parabola in the plane $((Eia+Eic)/Eid,(Eia-Eic)/Eid)$, 
it is natural for dispersions on individual veto channels to be non-gaussian.
\begin{figure*}[t] 
\centering 
\includegraphics[width=0.33\textwidth,keepaspectratio]{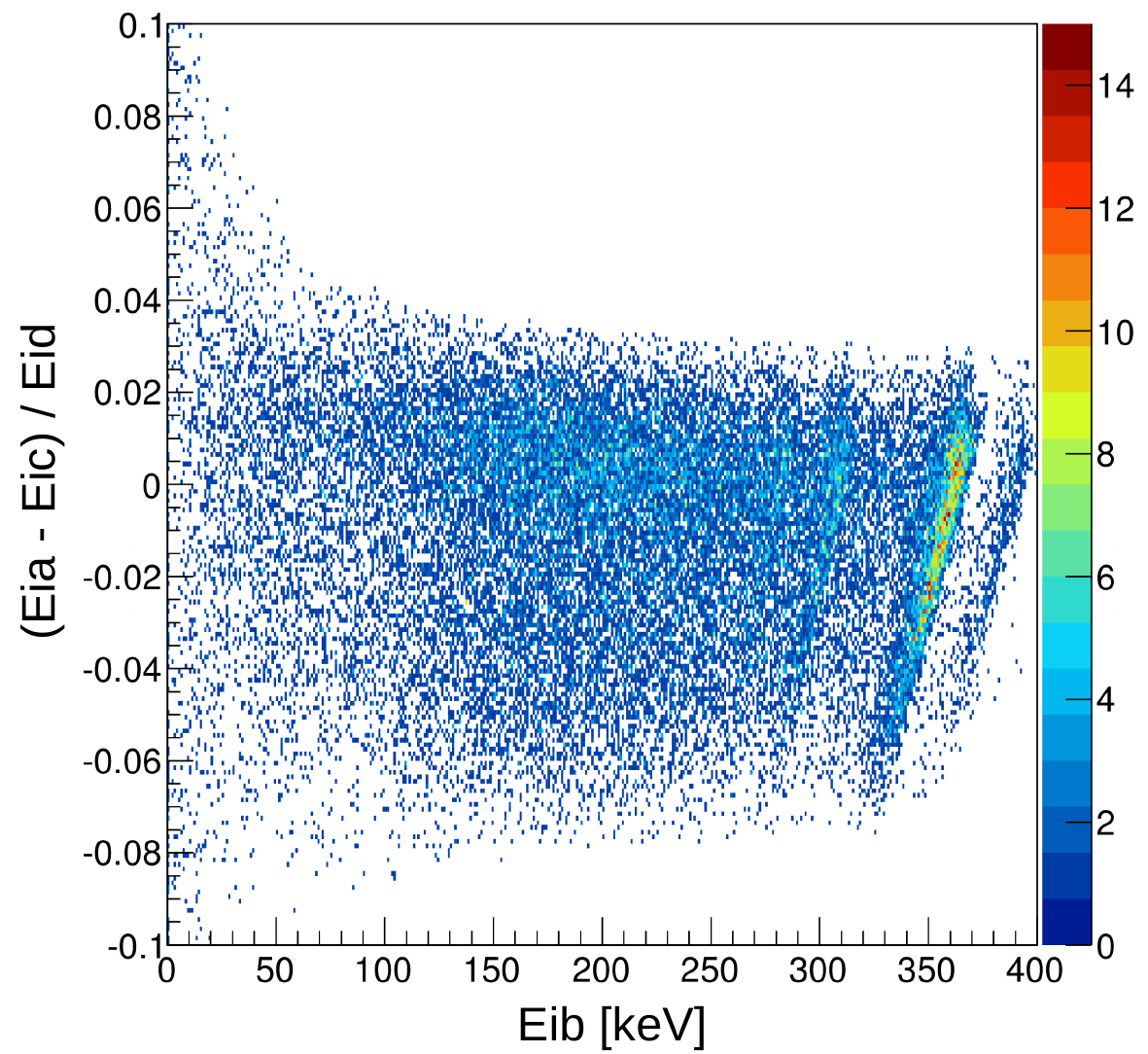}
\includegraphics[width=0.33\textwidth,keepaspectratio]{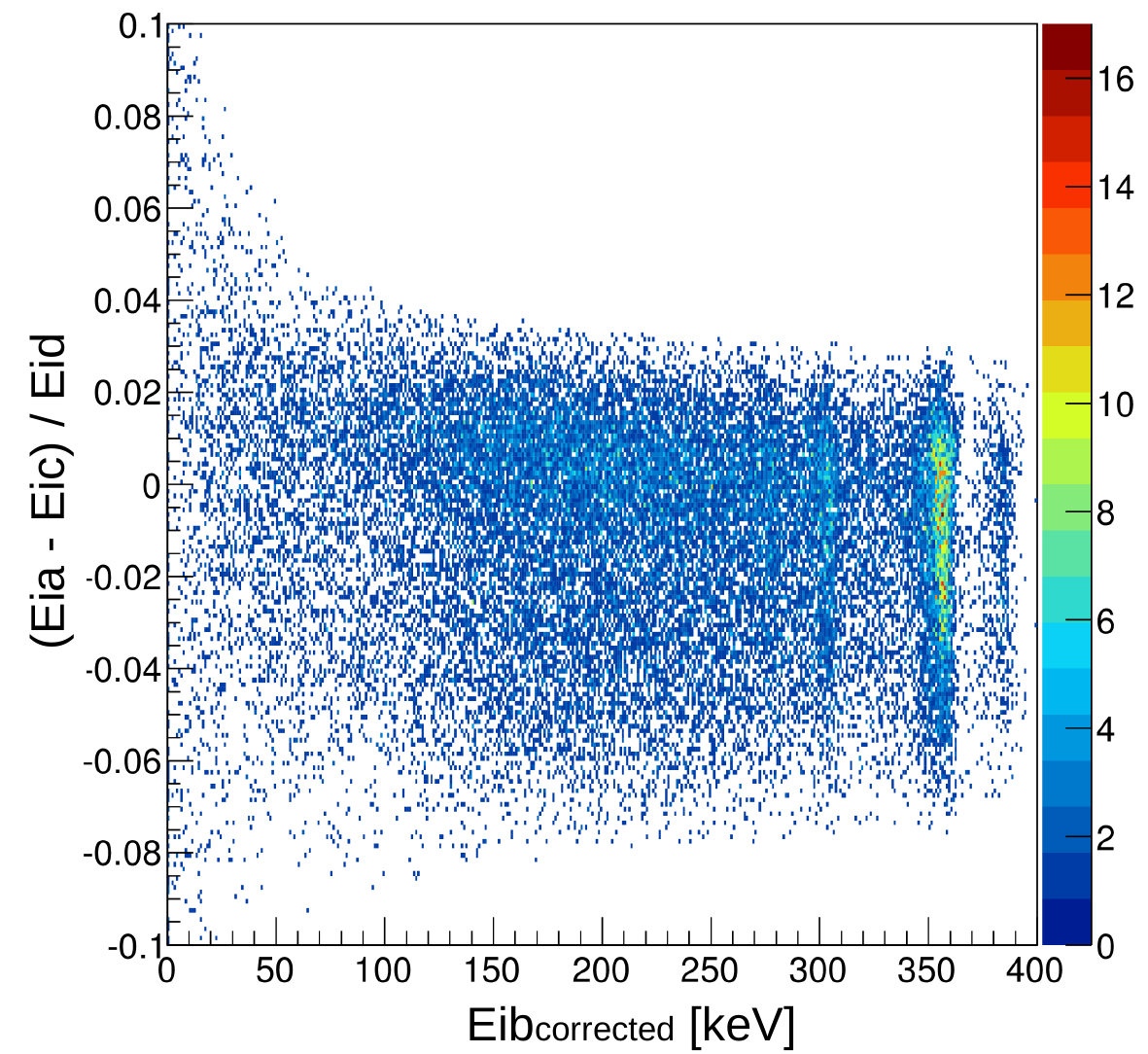}
\includegraphics[width=0.305\textwidth,keepaspectratio]{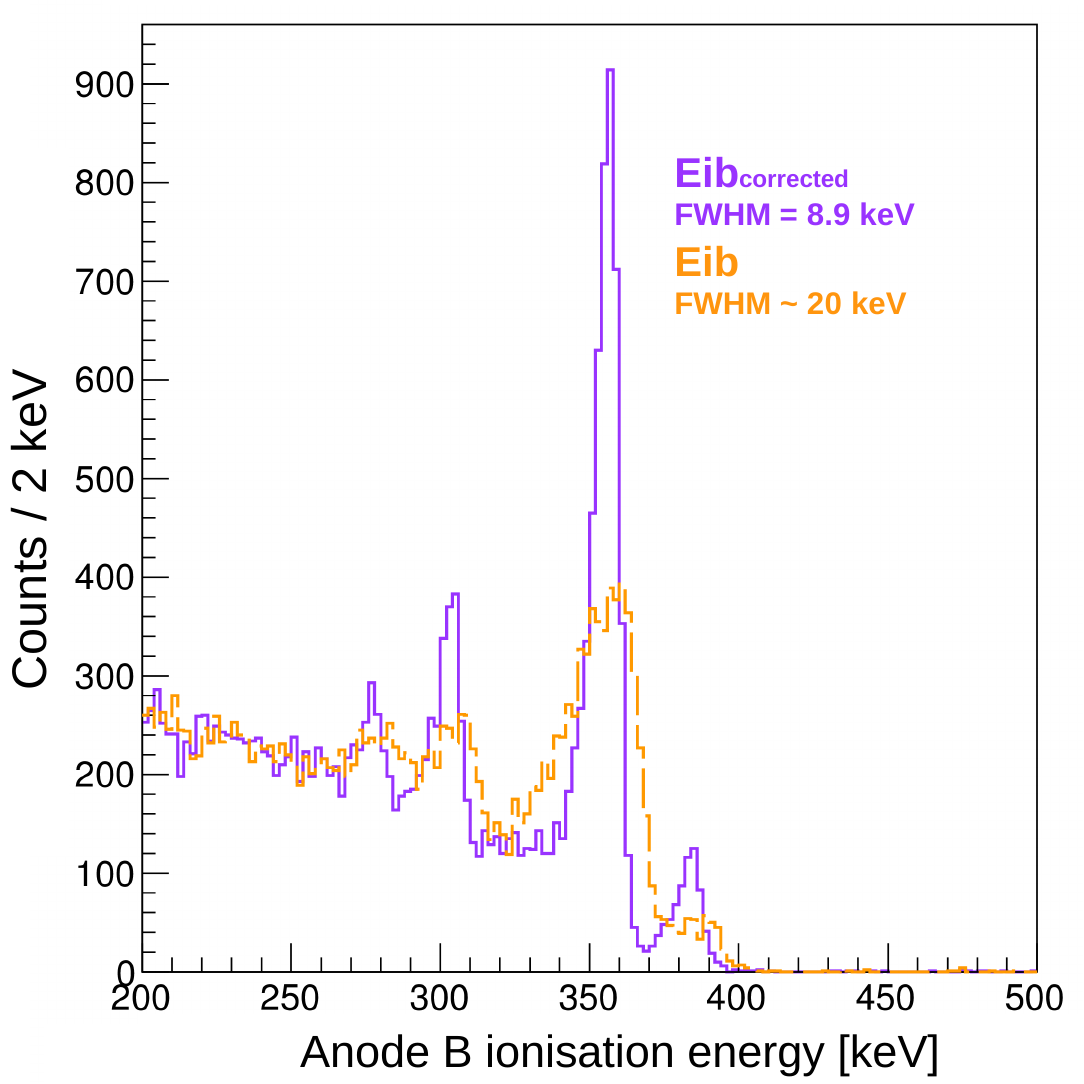}
\includegraphics[width=0.33\textwidth,keepaspectratio]{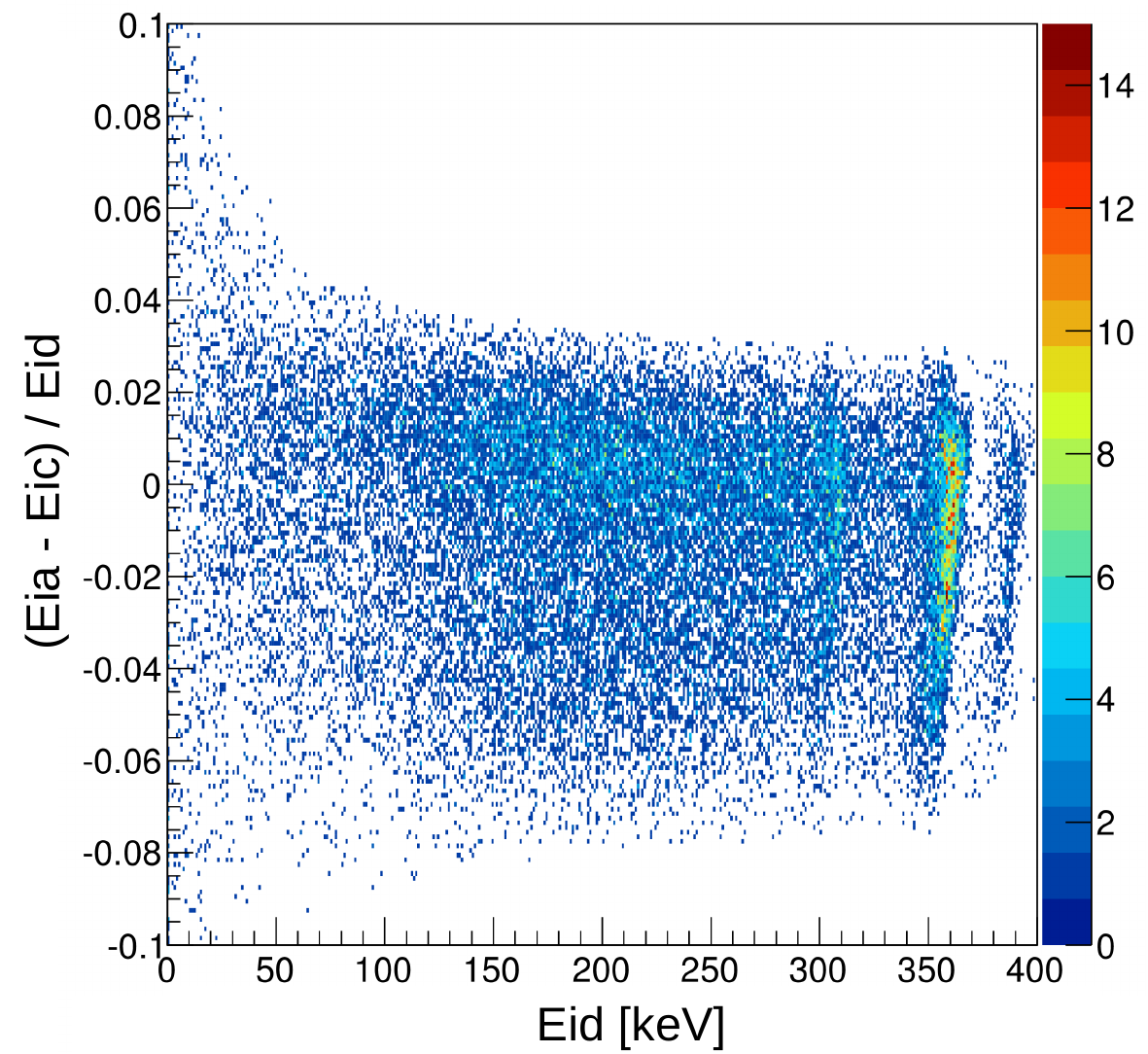}
\includegraphics[width=0.33\textwidth,keepaspectratio]{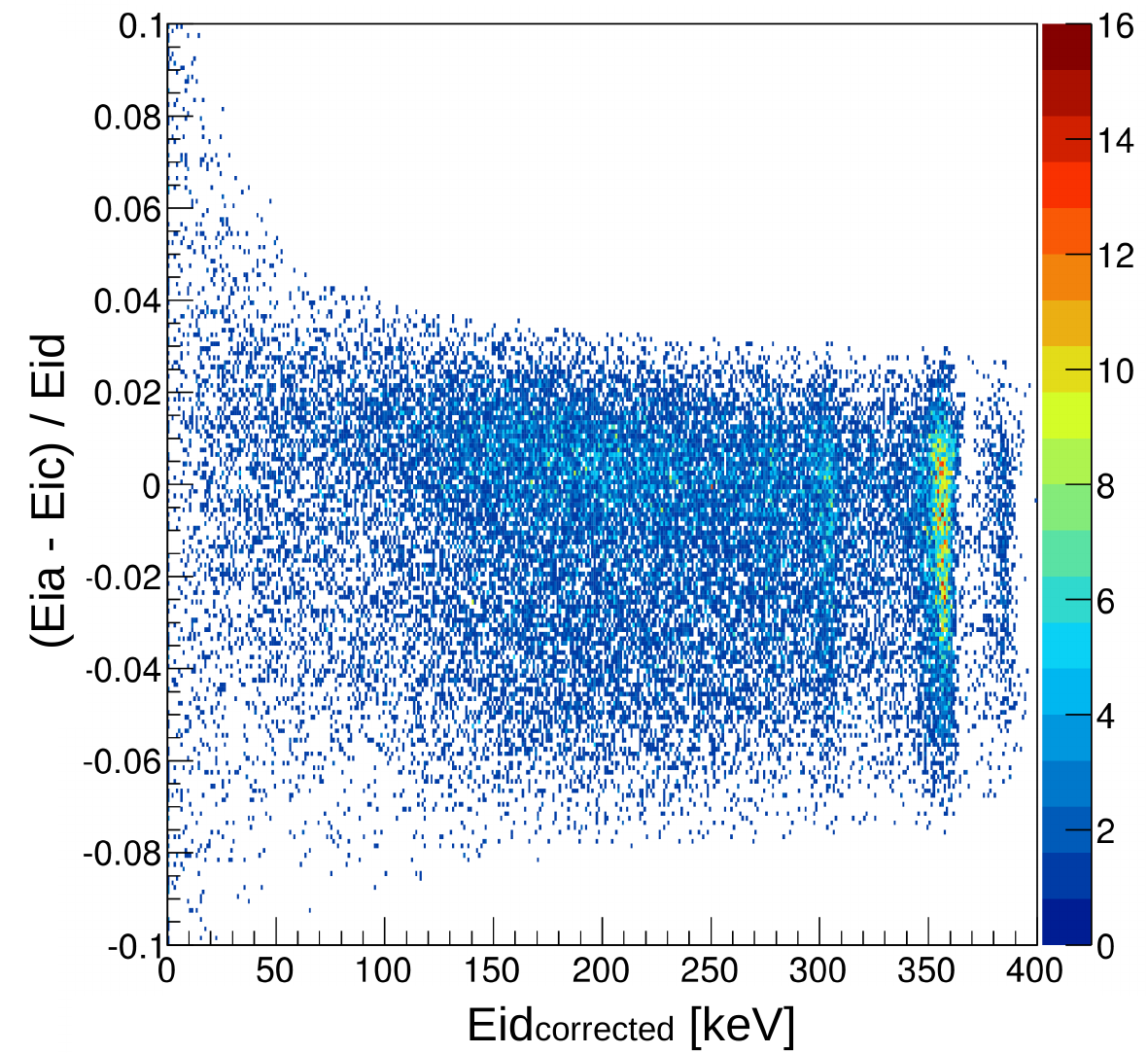}
\includegraphics[width=0.305\textwidth,keepaspectratio]{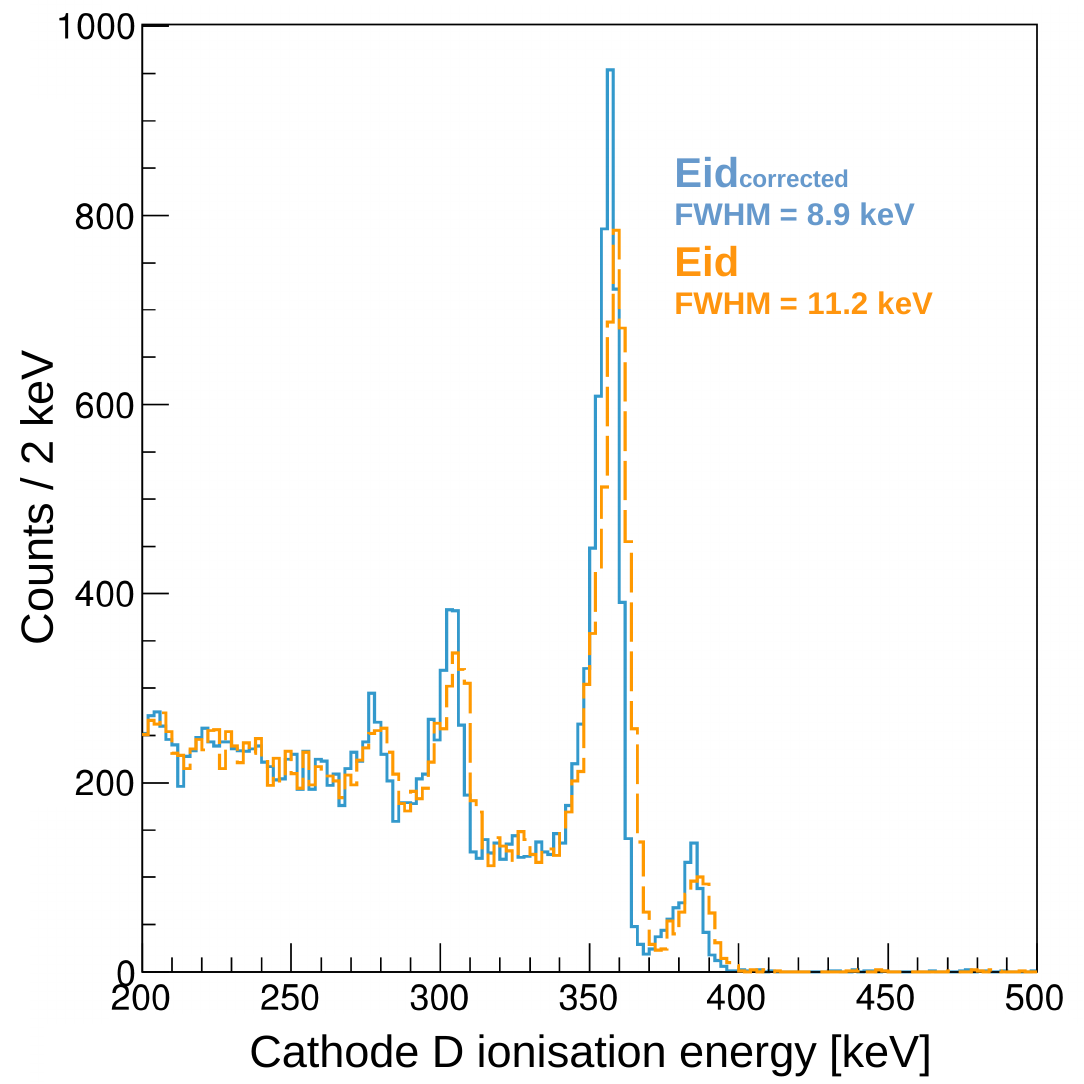}
\caption{\textit{Energy distribution of the fiducial anode B (top panels) and cathode D (bottom panels). Before correction, the deposit depth dependence of energy losses is visible via their correlation with $Z=(Eia-Eic)/Eid$ (left panels) and canceled once the correction is applied (middle panels). The resulting resolution improvement is shown on right panels with ionization energy spectra before (dashed lines) and after (solid lines) correction. A similar 8.9 keV resolution (FWHM) is reached on both channels. }}
 \label{figCORREC}
 \end{figure*}

\subsection{High energy corrections}

\label{sec:corrections}
Despite good individual baseline resolutions of 500 to $700\; \mathrm{eV}$ (FWHM) for most detectors, 
measured resolutions at 356 keV from  $^{133}\mathrm{Ba}$ ($\gamma$) calibrations 
are $\sim 10$ keV for the fiducial cathode D and $\sim 20$ keV for the fiducial anode B. 
According to our model, if the degradation of the resolution is originating from trapping, 
energy losses should be correlated to the energy deposit depth. 
As a matter of fact, such a dependence is clearly visible on left pannels of Fig.~\ref{figCORREC} through the correlation of the 356 keV line position with $(Eia-Eic)/Eid$ which was identified earlier to be our best estimate of the deposit depth in data. 
On all tested detectors, 
the dependence is always stronger for the fiducial anode (top panel) than for the fiducial cathode (bottom panel). 
Thus, we clearly identify the degradation of the resolution with energy as a consequence of charge carrier trapping in the bulk and highlight the fact that it is significantly larger for electrons than for holes. 
Moreover, we can use the 356 keV line to determine the deposit depth dependence of energy losses in order to correct trapping effects on the measurement. 
An empirical method consists at determining the 356 keV line position for different intervals of $Z=(Eia-Eic)/Eid$ and to fit this dependence with a smooth function $f(Z)$ such that the corrected energy measurement is given by $Eix_{corrected}=(356/f(Z))\times Eix$ where $x$ refers to the considered channel $x\in[b,d]$. 

We show on middle panels that the use of this corrected energy measurement straightens the distribution of events, 
which results in a consequent improvement of the resolution of both channels as shown on right panels: 
$\sim 20\%$ and $50\%$ for the cathodes B and D, respectively. 
The latter benefits much more from this correction as trapping is more important for electrons than holes.
After these corrections, the resolutions achieved are equivalent for both channels and equal to $8.9\;\mathrm{keV}$. 
Averaging both corrected energy measurements (as it is done to obtain $Efid$) does not lead to further improvement,
suggesting that the remaining energy dispersions for individual values of $Z$ are extremely correlated. 
This is certainly attributable to our inability to specifically select fiducial events of the near-homogeneous field region. 
The numerical model extended to the full detector volume indicates that shorter and longer drift distances associated to the field lines outside the central zone of the detector result in variations in the importance of trapping as the radius of the deposit extends into this region.

 \begin{figure*}[t]
\begin{center}
\includegraphics[width=0.49\textwidth,keepaspectratio]{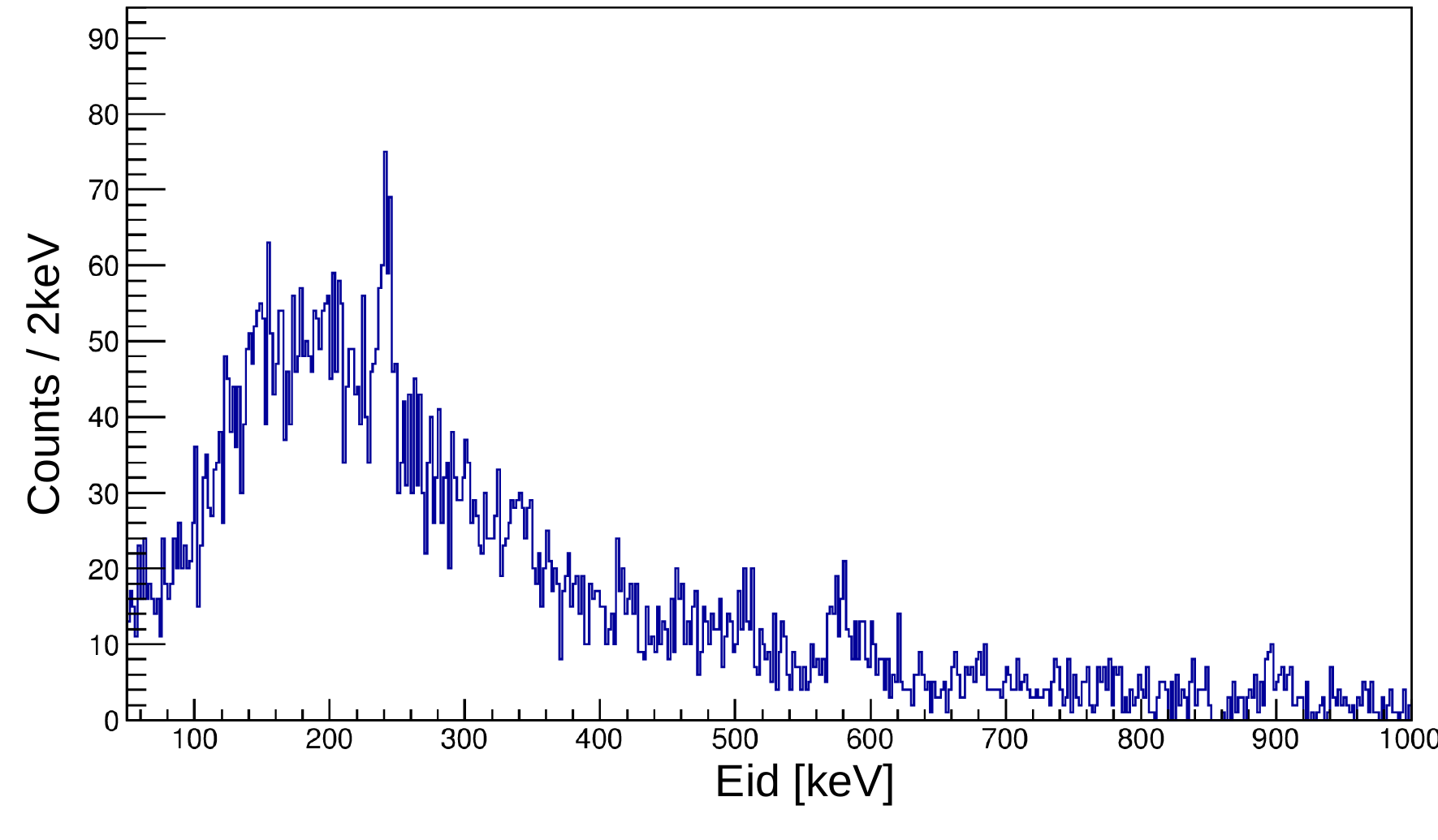} 
\includegraphics[width=0.49\textwidth,keepaspectratio]{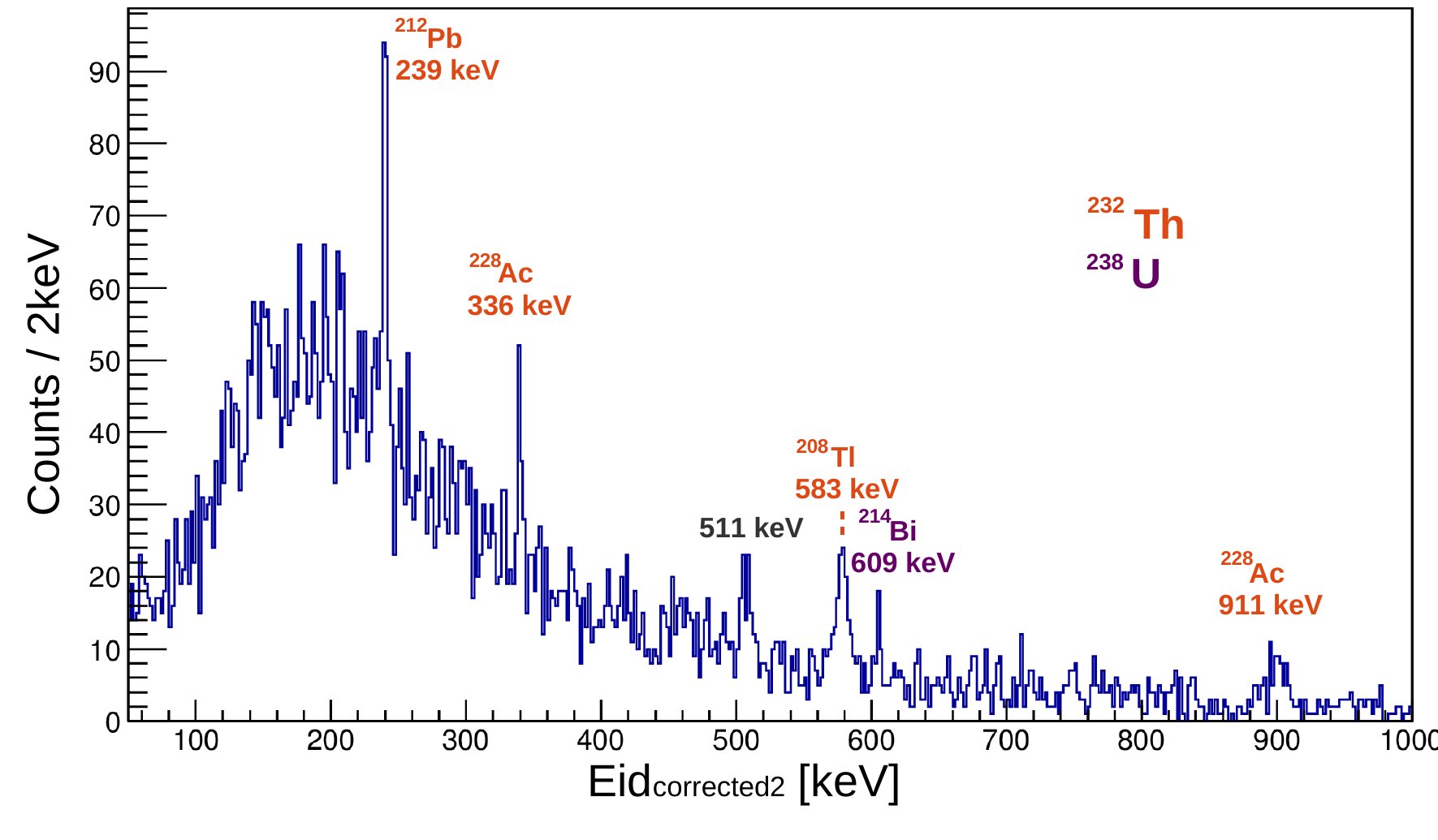} 
\caption{\textit{Ionization energy spectra of fiducial events recorded with the fiducial cathode D before (left) and after (right) correction measured with one detector in a WIMP search run. 
identified radiogenic lines  are shown on the right panel.}}
\label{correc1000}
\end{center}
\end{figure*}

A more detailed understanding of these effects would probably require to also take into account anisotropies in charge collection.
A practical constraint for deriving energy corrections is that the present procedure uses so far all of the independent linear
combination of the four electrode signals\footnote{We use $Eia-Eic$ to correct the fiducial energy $Eib+Eid$, and
$Eib-Eid$ = $Eia-Eic$ by charge conservation.}, except $(Eia+Eic)/Eid$.
Accordingly, an additional step has been introduced in the energy correction procedure 
in order to use the whole accessible trapping information contained in veto signals. 
Similarly to the first procedure discussed hereabove, we correct for the $Z=(Eia-Eic)/Eid$ dependence but 
only once we have used the remaining information contained in $R=(Eia+Eic)/Eid$ 
to reduce the energy dispersions for individual values of $Z$. 
Typically, we first define different intervals of $Z=\{Z_j\}$ and for each of these, 
we determine the value of $\alpha_j$ that provides the best resolution for the data subset $j$ 
with the estimator $Eix+R\alpha_jEix$. 
The different values of $\alpha_j$ as a function of $Z_j$ are fitted to a $g_{\alpha}(Z)$ second order polynomial. 
The correction to the whole data set is then given by:
\begin{equation}
Eix_{temp}=Eix+  R \times g_{\alpha}(Z) \times Eix
\label{temporary}
\end{equation}
To complete the procedure, we then proceed as before by fitting the dependence of $Eix_{temp}$ on $Z$ for 356 keV events with a smooth function $f(Z)$. 
The final correction is thus defined as:  
\begin{equation}
Eix_{corrected}=(356/f(Z))\times Eix_{temp}
\label{fullcorrection}
\end{equation}
The use of this full correction enables in average on tested detectors a resolution improvement of $\sim 65\%$ for the anode 
and $\sim 35\%$ for the cathode, allowing to reach a resolution of 7.3 and 7.1 keV, respectively. 
Here again, performances achieved with either B or D channel are similar and taking the average of the two corrected measurements does not improve the resolution. 
It was in fact to be expected due to charge conservation, as the correction combines three channels and consequently uses the whole accessible information since only three measurements are independent among the four. 
It is worth mentioning that the procedure also allows to improve the heat energy resolution by $\sim 30\%$. \\
Ionization and heat energy corrections start to be relevant above $50\; \mathrm{keV}$ only,
since at lower energies veto signals tend to be dominated by the baseline noise. 
The correction is particularly efficient for the detailed study of the gamma background, where the resolution for high-energy lines is instrumental for the identification of close-lying sources of radioactive isotopes. 
Figure \ref{correc1000} shows ionization energy spectra recorded in one EDELWEISS detector in WIMP search run, before (left panel) and after correction (right panel). 
Radiogenic 336 keV and 609 keV lines are clearly visible by the above described correction and allow to identify a nearby contamination by ${}^{228}\mathrm{Ac}$ and ${}^{214}\mathrm{Bi}$. 

\subsection{Improvements of the ionization baseline resolutions}

\label{sec:appchargeconserv} 
\begin{figure}[t]
\centering
\includegraphics[width=0.65\textwidth,keepaspectratio]{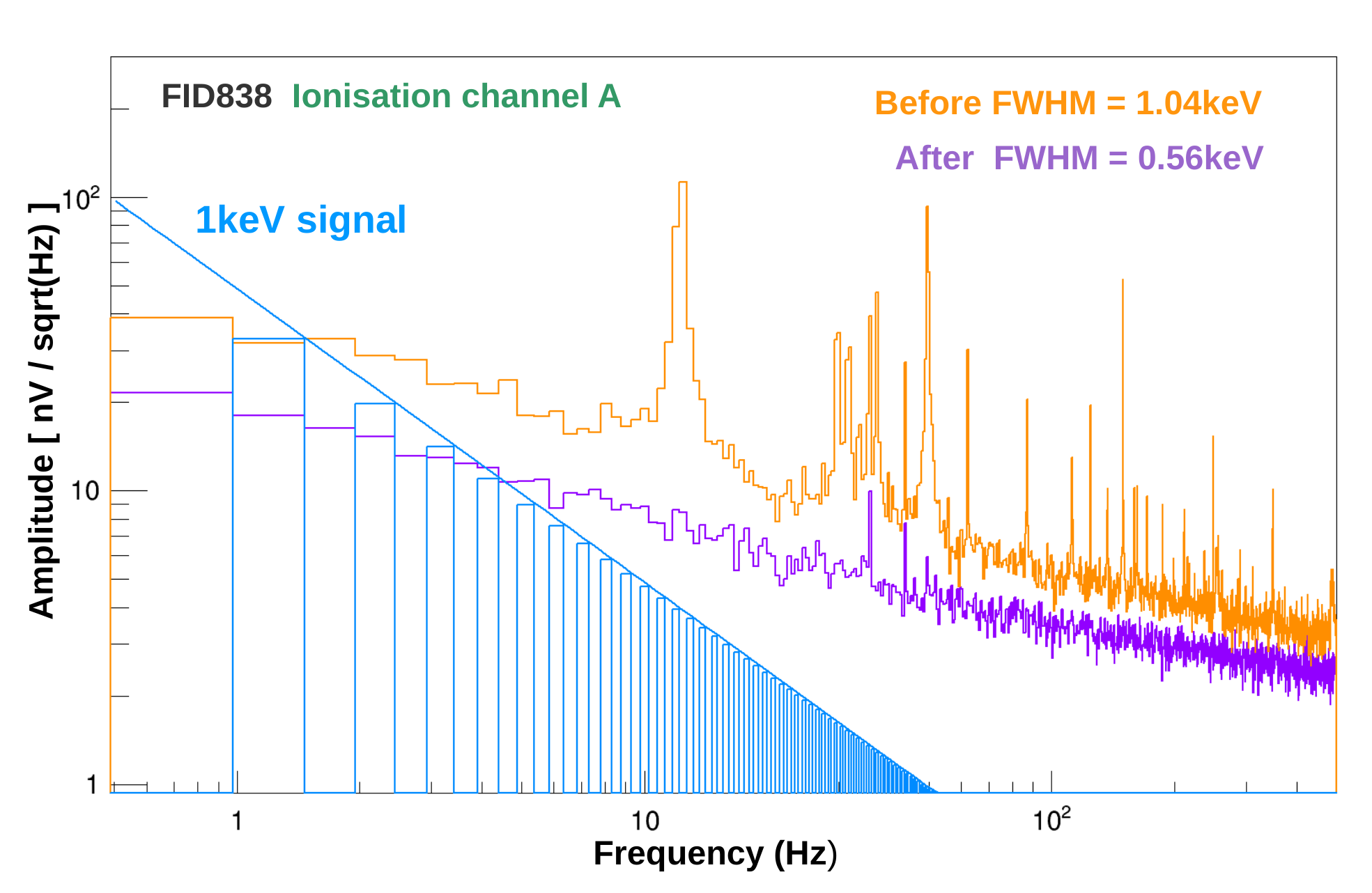}
\caption{\textit{Rms amplitude of the noise spectra of FID838 ionization channel A, before (orange histrogram) and after (purple histogram) taking advantage of charge conservation through Eq.~(\ref{secondform}). 
The blue histogram corresponds to the rms amplitude of a 1 keV signal (Heaviside in time domain), 
and the line is the envelop. 
Expected resolutions from an optimal filtering are shown on the picture.}} 
\label{fig:spectrum} 
\end{figure}

We previously mentioned a fundamental charge conservation relation (Eq.~(\ref{conservation})) independent of the trapping magnitude and location, due to a weighting potential associated to the sum of all the electrodes $\phi_{tot}=1$ in the whole detector. 
To be precise, this value varies from $\phi_{tot}=0.97$ in all of the bulk to $\phi_{tot}=1$ at the electrodes as FIDs do not act as perfect Faraday cages. 
The total net charge induced $Q_{tot}$ is therefore not exactly zero but given by the following relation: 
\begin{equation}
Q_{tot}=-e\left[(N_{Ce}-N_{Ch}) +(N_{Te}-N_{Th})\times0.97  \right] 
= 0.03\,e\,(N_{Te}-N_{Th})
\end{equation}
where $N_{Te}$ and $N_{Th}$ ($N_{Ce}$ and $N_{Ch}$) refer to the number of trapped (collected) electrons and holes,
respectively.
However, even in the extreme scenario in which one type of charge carrier would be significantly trapped, 
e.g. 20\% of the electrons, and the other type entirely collected, 
the total charge would be zero within 0.6\% of the total charge ($Q_{tot}=0.006eN_p$). 
Therefore, it is a reliable approximation to consider charge conservation (cf. Eq.~(\ref{conservation})) as always valid. 
The latter implies that the ionization measurement of any channel can either be directly obtained by reading it out or by using the three remaining channels. 
As a	 consequence, the most precise measurement of ionization channel A, for example,
 is not obtained by fitting a pulse template to the trace $A(t)$ but instead to $A^*(t)$ defined as:
\begin{equation}
A^*(t)=\alpha A(t)+(1-\alpha)(-B(t)-C(t)-D(t))
\end{equation} 
where the value of $\alpha$ that provides the best resolution $\sigma_{A^*}$ is the one that verifies $\partial \sigma_{A^*}/\partial \alpha=0$ and $\partial^2 \sigma_{A^*}/\partial \alpha ^2>0$. Let's consider the case in which all channels have the same baseline resolution $\sigma_0$ and the noise is uncorrelated. From quadratic error propagation, one gets that $\alpha=3/4$ minimizes $\sigma_{A^*}=[\alpha^2\sigma_0^2+3(1-\alpha)^2\sigma_0^2]^{1/2}$ and provides a resolution improvement such that $\sigma_{A^*}=(\sqrt{3}/2)\sigma_0$. This corresponds to:
\begin{eqnarray}
A^*(t)&=&\frac{3}{4}A(t)-\frac{B(t)+C(t)+D(t)}{4}\\
&=&A(t)-\frac{A(t)+B(t)+C(t)+D(t)}{4} \label{secondform}
\end{eqnarray}
where the latter expression highlights the fact that the correction to $A(t)$ does not affect the magnitude of the signal since the added term is zero irrespective of the origin of the signal. 
In \cite{benjaminschmidt}, this correction was introduced as a way to reduce correlated noise between the four electrodes. 
Here, it is shown that it  corresponds to an estimator that also minimizes uncorrelated errors. 
It also underlines that the correction does not affect the amplitude even in the case of charge trapping. 
Figure \ref{fig:spectrum} shows 
the rms noise amplitude $N(\nu)$ of channel A in the frequency domain for one detector, 
before and after application of Eq.~(\ref{secondform}). 
The rms amplitude $S(\nu)$ expected from a 1 keV signal is also shown. 
The correction is efficient at both reducing the peak structures due to correlated microphonics, 
as well as the smooth envelop, mostly due to uncorrelated noise. 
To assess the resolution improvement brought by this new energy estimator, we consider the expected resolution from the application of an optimal filter to the data:
\begin{equation}
\sigma=\left(\sum_{\nu} \frac{|S(\nu)|^2}{|N(\nu)|^2}\right)^{-1/2}
\end{equation}
We find that the use of the charge-conservation based correction provides a $\sim50$\% 
improvement of the resolution from $1.04\; \mathrm{keV_{ee}}$ to $0.56\; \mathrm{keV_{ee}}$ (FWHM). 
The additional gain in resolution relative to the expected gain of  15\% comes from the elimination
of correlated noise.
As $Efid$ is obtained via the traces $0.5(-B(t)+D(t))=0.5(-B^*(t)+D^*(t))$, 
the use of charge conservation does not improve the fiducial energy resolution. 
In both cases, the combination of both signals B and D result in a resolution of $\sigma_0/\sqrt{2}$ 
in the considered case where $\sigma_B=\sigma_D=\sigma_0$. 
However, the improvement on the individual electrodes A and C is extremely useful to increase the purity of the fiducial selection at low energy as it depends directly on the baseline resolution. 
This is particularly relevant in the context of low-mass WIMP search with EDELWEISS detectors \cite{EDWIIIlowmass}.

\section{Conclusion}

The analytical model for signals induced by bulk-trapped charges presented here considerably improves the understanding of  the functioning principle of FID detectors. 
It clearly establishes that charge carrier trapping in the fiducial volume is not only responsible for degrading the resolution of the fiducial electrodes but also for the large energy dependent dispersion of veto signals observed in data. 
These veto electrodes, until now exclusively used to reject surface events, 
allow for various other useful applications for WIMP searches. 
A statistical sensitivity to the energy deposit depth has been demonstrated from which we have derived an empirical method to correct both ionization and heat measurements and optimize fiducial background identification. 
Also, we have seen that some trapping information on ($\mu_e+\mu_h$) could be extracted from data even in absence of visible gamma-ray lines, 
providing a useful crystal quality criterium to select the detectors used for WIMP search. 
Finally, this study justifies the strategy of taking advantage of charge conservation in FID detectors to improve individual ionization baseline resolutions and consequently increase the surface event rejection power down to lower energies.

\section*{Acknowledgments}

The help of the technical staff of the Laboratoire Souterrain de Modane
and the participant laboratories is gratefully acknowledged. 
The EDELWEISS project is supported in part by the German ministry of science and education
(BMBF Verbundforschung ATP Proj.-Nr. 05A14VKA), 
by the Helmholtz Alliance for Astroparticle Physics (HAP), 
by the French Agence Nationale pour la Recherche and the LabEx Lyon Institute of Origins (ANR-10-LABX-0066) 
of the Universit\'e de Lyon in the framework Investissements d'Avenir (ANR-11-IDEX-00007), 
by the LabEx P2IO (ANR-10-LABX-0038) in the framework Investissements d'Avenir
(ANR-11-IDEX-0003-01) both managed by the French National Research Agency (ANR), 
by Science andTechnology Facilities Council (UK) and 
the Russian Foundation for Basic Research (grant No. 07-02-00355-a).

%\section*{References}

\bibliography{FIDbibliographie}

\end{document}